\documentclass[12pt]{article}

\usepackage{times}
\usepackage{graphics,graphicx,pdfpages}
\usepackage{flushend}
\usepackage{hyperref}
\usepackage{caption}
\usepackage{subfigure}
\usepackage{amssymb}
\usepackage{geometry}
\usepackage{ragged2e}
\usepackage{flushend}
\usepackage[ruled,linesnumbered,vlined]{algorithm2e}  
\usepackage{amsmath}  
\usepackage{bm} 
\usepackage{multirow}
\usepackage{lscape}
\usepackage{cite}
\usepackage{ragged2e}

\makeatletter
\newcommand{\rmnum}[1]{\romannumeral #1}
\newcommand{\Rmnum}[1]{\expandafter\@slowromancap\romannumeral #1@}
\makeatother

\title{Massive Self-Assembly in Grid Environments} 

\author
{Wenjie Chu, Wei Zhang,$^\ast$ Haiyan Zhao, Zhi Jin,$^\ast$
Hong Mei\\
\\
\normalsize{ Department of Computer Science and Technology, Peking University, China}\\
\normalsize{Key Laboratory of High Confidence Software Technology (Peking University), MoE of China}\\
\normalsize{$^\ast$Correspondence to: (zhangw.sei, zhijin)@pku.edu.cn}
}

\begin{document} 
\baselineskip22pt
\maketitle 

\begin{abstract}
  Self-assembly plays an essential role in many natural processes, involving the formation and evolution of living or non-living structures, and shows potential applications in many emerging domains. In existing research and practice, there still lacks an ideal self-assembly mechanism that manifests \emph{efficiency}, \emph{scalability}, and \emph{stability} at the same time. Inspired by \emph{phototaxis} observed in nature, we propose a computational approach for massive self-assembly of connected shapes in grid environments. The key component of this approach is an \emph{artificial light field} superimposed on a grid environment, which is determined by the positions of all agents and at the same time drives all agents to change their positions, forming a dynamic mutual feedback process. This work advances the understanding and potential applications of self-assembly.
\end{abstract}

\section{Introduction}
As a kind of interesting and mysterious phenomenon, self-assembly has been unintentionally observed in many natural processes and often appears in science fiction movies. In the 2014 animated movie ``Big Hero 6'', one of the impressing scenes involves thousands of micro-robots assembling themselves into arbitrary shapes and transforming between shapes dynamically, which shows pervasive potential applications of self-assembly from the perspective of imagination. Before being perceived by human beings, various kinds of self-assembly phenomenon have existed in nature for a long time \cite{Whi02}, playing essential roles in the forming of multi-component non-living structures \cite{Grzybowski09, Estroff04, Marsh15}, multi-cellular living organisms \cite{Grzybowski09, Weijer09, Mehes14}, and multi-organism biological systems \cite{camazine03, mlot11}.
These self-assembly phenomena, either real or fictional, all implicitly point to an important research problem: \emph{whether we can construct artificial self-assembly systems}. The benefit of resolving this problem is twofold: on the one hand, it would contribute to a deep understanding of self-assembly mechanisms; on the other hand, it would facilitate the applying of self-assembly in many valuable scenarios, including autonomous cooperation of UAVs \cite{Finn07, KevinZ18} and intelligent transportation systems \cite{Viriyasitavat12,Daniel18}.

The problem of constructing artificial self-assembly systems has attracted increasing attention in recent years, but there still lacks an ideal self-assembly mechanism that manifests the three essential features of \emph{efficiency}, \emph{scalability}, and \emph{stability} at the same time.
Edge-following based methods \cite{rubenstein14,tucci18} lead to a self-assembling process with \emph{low efficiency}, because of the heavily-decreased degree of parallelism. 
Path planning/scheduling methods based on prior task allocation \cite{JYu13,alonso11} suffer from \emph{poor scalability} concerning the number of agents involved in self-assembly due to the high computational cost of global shortest path generation and task assignment.
Methods based on \emph{artificial potential fields} (APF) \cite{sabattini09,chiang15,falomir18,bi18,wolf04} behave well in efficiency and scalability, but make a poor showing in \emph{stability}, because agents may be trapped in \emph{local minima}, which will be more likely to appear as the number of agents increases \cite{gayle09,chiang15}.
Although some improved APF-based methods \cite{falomir18, sabattini11} have been proposed to eliminate local minima, they can only cope with limited scenarios.

In general, self-assembly can be viewed as a kind of collective intelligence (CI) phenomena: a group of agents with limited capabilities exhibits collective intelligent behavior that goes well beyond individual capabilities. 
Existing research offers two complementary understandings of CI:
in the \emph{explanatory} understanding \cite{theraulaz99}, a key element in CI is the \emph{environment}, which acts as an external memory of a collective of agents and drives each agent's behavior based on the information the agent perceives from the current environment; 
in the \emph{constructive} understanding \cite{zhang20}, the key to build a problem-oriented artificial CI system is to enable and maintain an ongoing loop of \emph{information exploration}, \emph{integration}, and \emph{feedback} among agents in the collective, until an acceptable solution to the target problem emerges.

Guided by the two understandings of CI, we propose here a computational approach for massive self-assembly of connected shapes in grid environments. This approach mimics the \emph{phototaxis} observed in many species \cite{Jekely2009} (i.e., organisms' movement towards or away from light sources), by superimposing the grid environment with an \emph{artificial light field} (ALF), which plays the dual role of an \emph{external memory} of self-assembling agents in the explanatory CI understanding and a \emph{carrier} for information integration and feedback in the constructive CI understanding. The essence of this approach is a mutual feedback process between the ALF and the agent collective: the current positions of all agents determine the current state of the ALF, which in turn drives the agents to further change their current positions. In experiments, this approach exhibits high efficiency, scalability and stability in a set of diverse shape formation tasks.
In an extreme case involving 5469 agents in a 135$\times$135 grid environment, this approach accomplishes the self-assembly task accurately with only 119 steps/256.6 seconds on average. Compared to the state-of-the-art centralized distance-optimal algorithm, this approach exhibits a $n^3$ to $n^2\mathit{log}\left(n\right)$ decrease in the absolute completion time of self-assembly tasks with respect to task scale $n$, and can be easily accelerated through parallelization.

\section{Method}
The proposed approach consists of five components (Figure 1.A): a grid environment $G$, a target shape $S$, an agent collective $A$, an artificial light field $\mathcal{F}$ superimposed on $G$, and a lightweight coordinator $C$. 
$C$ and all agents in $A$ form a star topology: each agent connects with $C$ through a communication channel; no communication channel exists between any two agents. $C$ coordinates each agent's behavior by playing three roles: a \emph{generator} of discrete system times, a \emph{recorder} of system states, and an \emph{actuator} of grid locking/unlocking. When resolving a self-assembly problem, each agent interacts with $C$ through an iterative process (Figure 1.B). 
Before the process begins, each agent $a_i$ reports its initial position $p_0(a_i)$ to $C$; as a result, $C$ gets the system state at time $0$, denoted as $p_0$. After that, $C$ broadcasts $p_0$ to each agent in $A$.
In every iteration, each agent sequentially carries out three actions:
(\rmnum{1}) \emph{local ALF calculation}, where the agent retrieves other agent's position from the $C$, identify light sources and calculate its local ALF;
(\rmnum{2}) \emph{priority queue generation}, where the agent constructs a priority queue of next positions based on the agent's state and local ALF;
(\rmnum{3}) \emph{next position decision}, where the agent cooperates with $C$ to obtain a conflict-free next position.
After that, the agent will move to its next position, inform $C$ this movement, and enter the next iteration.
The process will terminate when agents form the target shape.
(See supplementary for more details)

\begin{figure*}[t]
    \centering
    \includegraphics[scale=0.55,trim={0cm 0cm 0cm 0cm},clip]{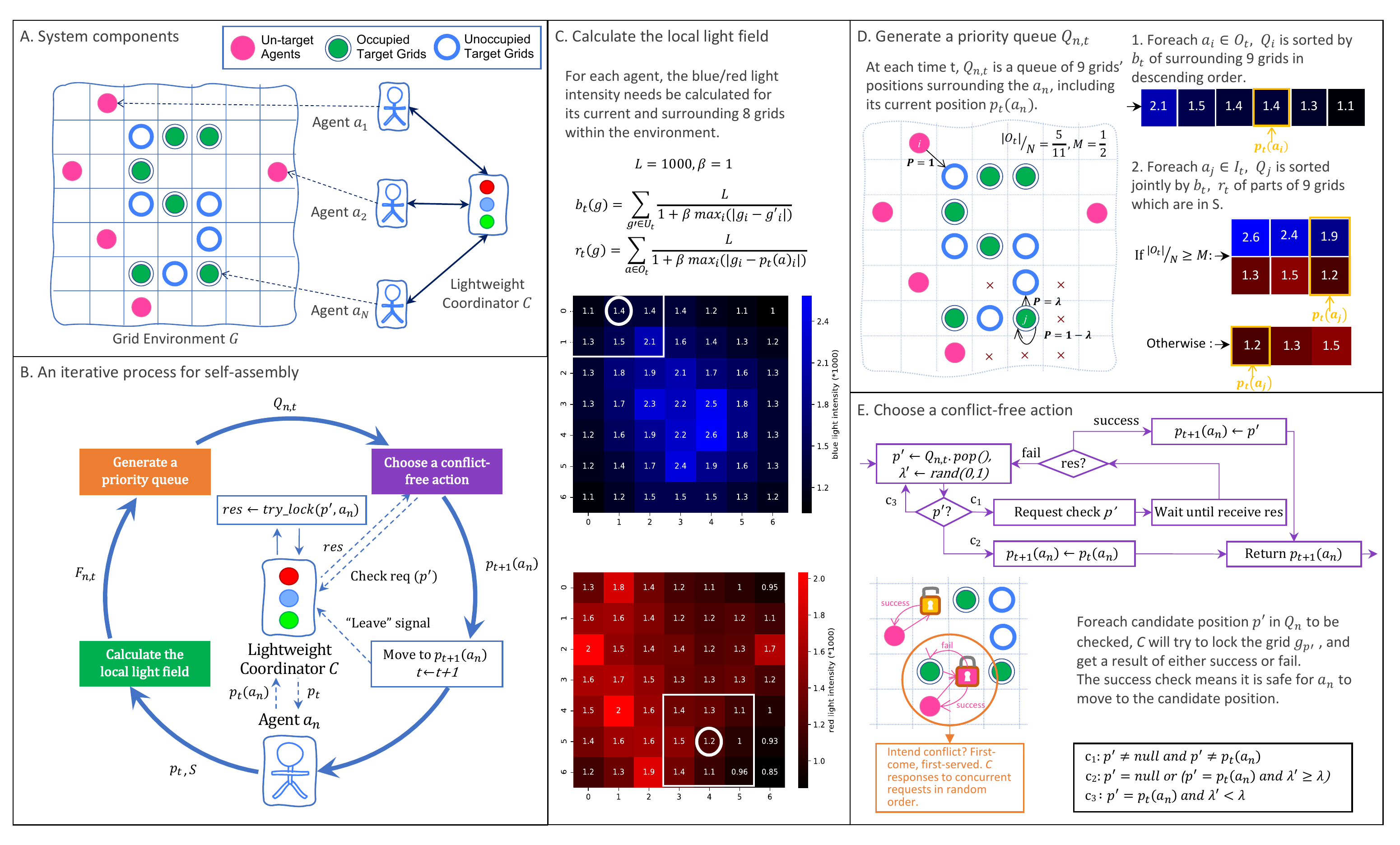}
    \caption{An ALF-based self-assembly system. A. The system's main components; B. An iterative process for self-assembly; C. Method to calculate the local light field for each agent; D. Policy to generate the priority queue of next positions for each agent; E. The decision process between an agent and the lightweight coordinator to choose a conflict-free next action.}
    \label{fig:arch}
\end{figure*}

\textbf{Local ALF calculation.} Each agent calculates its local ALF, i.e., the light intensities in its surrounding 8 grids as well as its current position, based on the system state at current time $t$, namely $p_t$. 
The ALF at time $t$, denoted as $\mathcal{F}_t$, is defined as a pair of functions $(r_t, b_t)$, where the former maps each grid to the intensity of red light at the grid, and the latter to the intensity of blue light. The intensity of red/blue light at a grid is the sum of all red/blue light sources' intensities at the grid.
At any time $t$, each agent out of the shape is a source of red light, and each unoccupied target grid is a source of blue light. The intensity of the light attenuates with propagation distance. 
As a result, given a $g \in G$, $r_t(g)$ and $b_t(g)$ can be defined conceptually as follows:
$r_t(g) = \sum_{a \in O_t} f(L, \alpha, dis(g, p_t(a)))$, 
$b_t(g) = \sum_{g' \in U_t} f(L, \alpha, dis(g, g'))$,
where $O_t$ is the set of agents out of the shape at $t$, $U_t$ is the set of unoccupied target grids at $t$, $L$ is the intensity of light emitted by a light source, $\alpha$ is the attenuating rate of light, $dis$ is a function that returns the distance between two grids, and $f$ is a function that returns the intensity of light after the light has traveled a certain distance from its source.

\textbf{Priority queue generation.} Given an agent $a_n$ at time $t$, its priority queue of next positions, denoted as $Q_{n,t}$, is a permutation of $a_n$'s local 9 grids, generated based on its local $\mathcal{F}_t$. The strategy for generating $Q_{n,t}$ depends on $a_n$'s state. When $a_n$ is outside the target shape, the local 9 grids are sorted in descending order by blue light intensity; this strategy directs agents outside the target shape to move towards the shape. When $a_n$ is already inside the target shape, $Q_{n,t}$ will be constructed according to parameter $\omega$, the ratio of agents outside the target shape to all agents: when $\omega>$ 0.15, the local 9 grids are sorted in descending order by blue light intensity and ascending order by red light intensity; when $\omega\le$ 0.15, the local 9 grids are sorted in ascending order by red light intensity. The former strategy motivates an agent to keep moving towards those unoccupied positions in the center of the target shape after the agent has entered the shape, and the latter one motivates an agent to leave the peripheral positions of the target shape.

\textbf{Next position decision.} After obtaining $Q_{n,t}$, agent $a_n$ will cooperate with $C$ to decide a conflict-free next position from $Q_{n,t}$, through an iterative decision process. In each iteration, agent $a_n$ first retrieves the head element of $Q_{n,t}$, denoted as $\hbar$. If $\hbar=p_t\left(a_n\right)$, $a_n$ will immediately go to the next iteration with the probability of $\gamma$; otherwise, $a_n$ will send $\hbar$ to $C$ to check whether $\hbar$ is conflict-free or not. If $\hbar$ is conflict-free, $a_n$ will use $\hbar$ as its next position and terminate the decision process; otherwise, $a_n$ will go to the next iteration. In the extreme case when $Q_{n,t}$ becomes empty and the decision process has not terminated, $a_n$ will use $p_t\left(a_n\right)$ as its next position and terminate the process. $C$ uses a try\_lock mechanism to determine whether $\hbar$ is conflict-free for $a_n$ or not; each grid in the environment is treated as a mutex lock \cite{Dalessandro11}. When receiving $\hbar$ from $a_n$, C will try to acquire the lock of $\hbar$ for $a_n$: if $\hbar$'s lock is not held by any other agent, $C$ will assign the lock to $a_n$ and return a success signal; otherwise, a \emph{fail} signal will be returned.

\section{Results}
To evaluate the effectiveness of this approach, we conducted a set of experiments, involving 156 shapes from 16 categories (See supplementary for more details). Four methods are selected as baseline:
(1) OPT-D \cite{JYu13}, a centralized distance-optimal method for self-assembly;
(2) HUN \cite{alonso11}, an iterative self-assembly method based on global task allocation with Hungarian algorithm;
(3) DUD \cite{bi18}, a self-assembly method based on artificial potential field; 
(4) E-F \cite{rubenstein14}, a gradient-based edge-following method for self-assembly.
In particular, we focus on three measures of \emph{completion quality} $\rho$, \emph{relative completion time} $t$, and \emph{absolute completion time} $\tau$:
$\rho$ denotes the \emph{shape completion degree} of the agent swarm when achieving a stable state; $t$ denotes \emph{the number of iterations} to complete a target shape; and $\tau$ denotes the \emph{physical time} to complete a target shape.
The three measures are analyzed from three aspects: \emph{efficiency}, \emph{scalability}, and \emph{stability}.
Parts of the experiments are shown in Figure 2 and Movies S1-S3.

\begin{figure*}[t]
    \centering
    \includegraphics[scale=0.16,trim={0cm 0cm 0cm 0cm},clip]{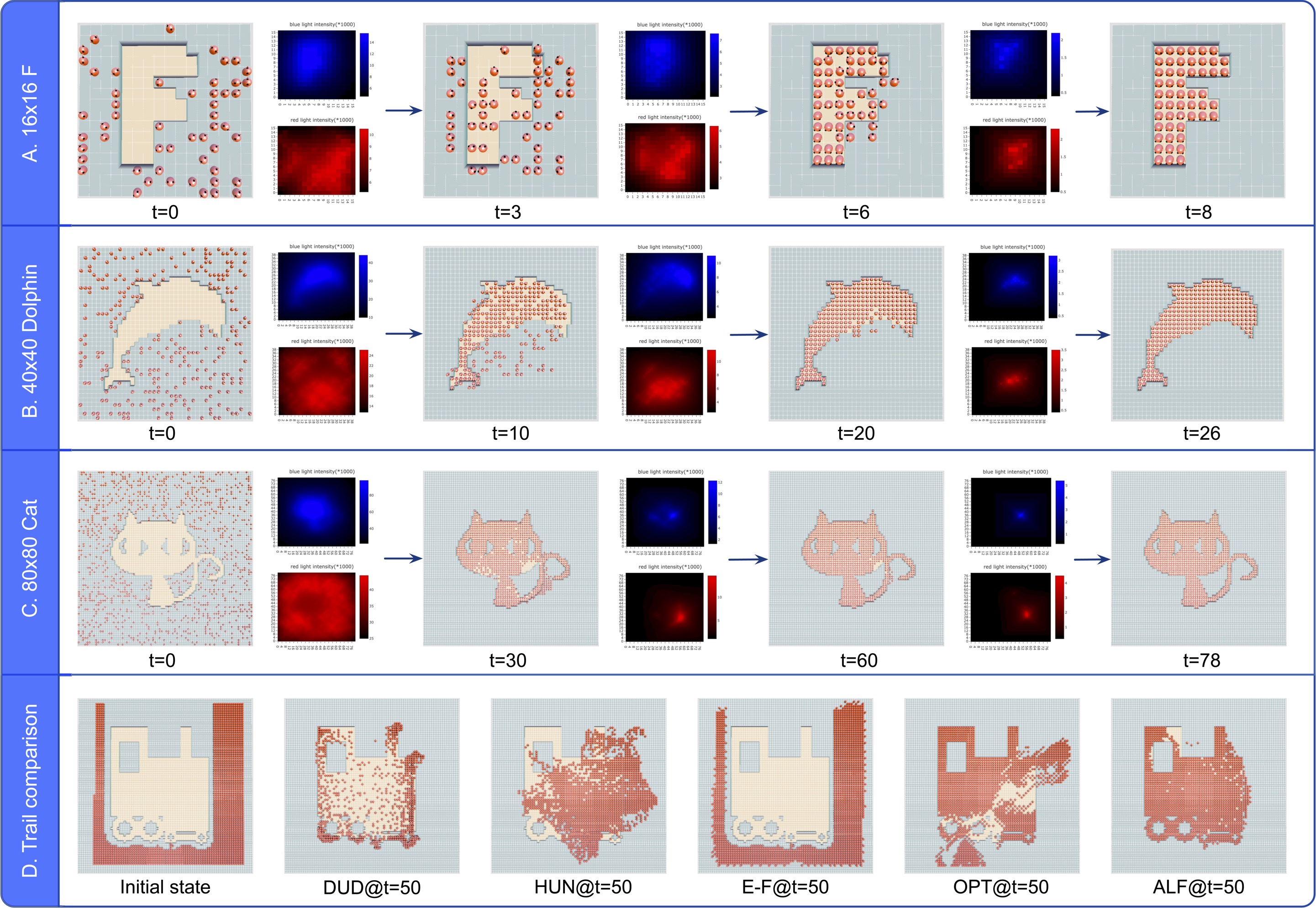}
    \caption{Trails when forming different shapes in grid environments with different scales. A. System states at four time steps ($t=$ 0, 3, 6, 8) when forming letter ``F'' in a 16$\times$16 grid environment with 52 agents, using ALF; B. System states at four time steps ($t=$ 0, 10, 20, 26) when forming shape ``dolphin'' in a 40$\times$40 grid environment with 276 agents, using ALF; C. System states at four time steps ($t=$ 0, 30, 60, 78) when forming shape ``cat'', which has inner holes, in a 80$\times$80 grid environment with 1033 agents, using ALF; D. System states at $t=$ 50 when forming shape ``locomotive'' in a 80$\times$80 grid environment with 1785 agents and with a fixed initial state (i.e., state at $t=$ 0), using five different methods of DUD, HUN, E-F, OPT-D, and ALF (the proposed approach).}
    \label{fig:trail}
\end{figure*}

\textbf{Efficiency}. To evaluate the efficiency, we compare our approach with the baseline methods under two different policies of agents' initial distribution:
\emph{random} and \emph{specific} policies. 
Experiments with the \emph{random} policy are carried out in three environments with scale of 16$\times$16, 40$\times$40, and 80$\times$80, respectively.
In each environment, for each of the 156 shapes, we observe the three measures of $\rho$, $t$ and $\tau$ of different methods when resolving the same self-assembly problem.
Figure 3.A shows the experimental results of a randomly-selected shape (\emph{locomotive}) in 80$\times$80 environment.
Table S2 and S3 gives each method's performance on 16 representative shapes (listed in Table S1) and 16 categories of shapes, respectively. 
It is observed that:
(1) for completion quality, our approach shows nearly the same performance ($99.9\%$) with OPT-D ($100\%$) and outperforms HUN ($\times 1.094$) and DUD ($\times 1.126$) on all 156$\times$3 shapes.
(2) for relative completion time, our approach performs worse than OPT-D ($\times 3.032$), and better than HUN ($\times 0.074$) and DUD (which fails in all 50 repeated experiments) on all 156$\times$3 shapes ; 
(3) for absolute completion time, our approach outperforms OPT-D ($\times 0.007$) and HUN ($\times 0.173$).
E-F method is not included in the comparison, because of its specific requirement on agents initial distribution.

Experiments with the \emph{specific} policy are carried out in the 80$\times$80 environment for the 16 shapes listed in Table S1, and evaluated by the same measures with random-policy experiments.
Figure 3.B illustrates the experimental results for a shape (\emph{locomotive}); see table S4 for complete results. 
It is observed that: 
(1) for completion quality, our approach shows nearly the same performance ($99.9\%$) with OPT-D ($100\%$), and outperforms the HUN ($\times1.101$), DUD ($\times1.068$), and E-F ($\times1.352$) on all 16 shapes;
(2) for relative completion time, our approach performs worse than OPT-D ($\times 1.105$), and better than HUN/DUD (which fails in all 20 repeated experiments) and E-F ($\times 0.017$);
(3) for absolute completion time, our approach outperforms OPT-D ($\times 0.018$) and E-F ($\times 0.040$);
(4) the E-F method, using the edge-following strategy, shows the longest/second-longest relative/absolute completion time (e.g., in the \emph{r-6-edge} task with 1595 agents, this method takes 6156 iterations/703.9 seconds, while our approach only 75 iterations/16.7 seconds);

\textbf{Scalability}.
To evaluate the scalability, we compare our approach with OPT-D on both $t$ and $\tau$ for 16 shapes (listed in Table.S1) in 12 different shape scales.
Figure 3.C and 3.D shows the experiment results of shape ``irre-curve-1'' on $t$ and $\tau$, respectively; see Figure S3-S4 for results of all the 16 shapes.
It is observed that: 
(1) for relative completion time $t$, both our approach and OPT-D shows a $log(n)$ increasing as the shape scale $n$ grows ($R^2=0.9506$ and $0.9799$, respectively);
(2) for absolute convergence time $\tau$, our approach shows a $n^2log(n)$ increasing ($R^2=0.9829$), while OPT-D a $n^3$ increasing ($R^2=0.9996$).
In addition, our approach is easy to parallelize, 
e.g., with 16 threads, our approach achieves an average \emph{parallel speedup} of 12.96, leading to a 92.28\% decreasing of $\tau$.

\textbf{Stability}.
To evaluate the stability, we analyze the standard deviations of $\rho$, $t$, and $\tau$ of our approach on 50 randomly-initialized experiments for each of the 156 shapes in each of the three environments with scale of 16$\times$16, 40$\times$40, and 80$\times$80, respectively. Table S2 and S3 gives the complete results. 
It is observed that our approach shows a normalized $\sigma(\rho)$, $\sigma(t)$ and $\sigma(\tau)$ of 0.00034, 0.04410, and 0.00033, respectively;

\begin{figure*}[t]
    \centering
    \includegraphics[scale=0.28,trim={0cm 0cm 0cm 0cm},clip]{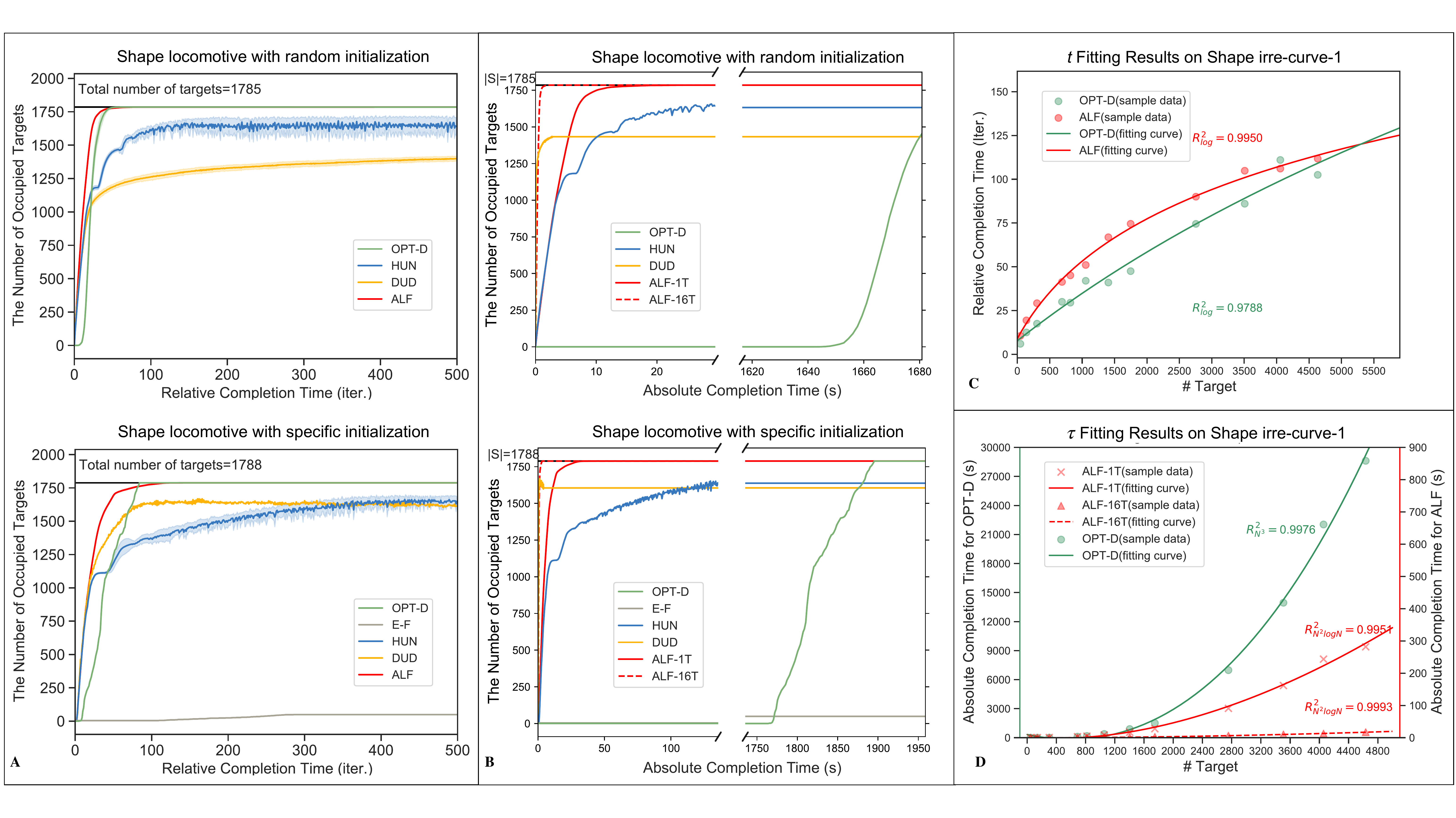}
    \caption{The statistical results of different methods' efficiency and scalability. A. In an 80$\times$80 environment with random and specific initialization, respectively, the task progress of different methods when forming shape ``locomotive'' as the relative completion time increases; B. In an 80$\times$80 environment with random and specific initialization, respectively, the task progress of different methods when forming shape ``locomotive'' as the absolute completion time increases; C. The changing trend of relative completion time by ALF and OPT-D as the number of targets increases; D. The changing trend of absolute completion time by ALF and OPT-D as the number of targets increases; in addition to the 1-thread ALF (ALF-1T, i.e., the ALF approach running in a 1-thread hardware environment), the 16-thread ALF (ALF-16T) is also investigated.}
    \label{fig:statis-3}
\end{figure*}

In addition, we also observe that our approach shows a hole-independent property (i.e., the existence of holes in a shape does not affect the performance of an approach), which is missing in many existing methods \cite{rubenstein14, tucci18}.

In nature, self-assembly phenomena emerge from collective behaviors of swarms based on chemical or physical signals, whereas in our approach, we designed a kind of digital signals, namely \emph{artificial light field}, to enable a massive swarm of agents to gain such ability in grid environments. Experiments have demonstrated the superiority of our approach in constructing massive self-assembly systems: for a self-assembly task with $n$ agents, the absolute completion time of our approach is decreased from the magnitude of $n^3$ to $n^2log\left(n\right)$, and can be further decreased through parallelization. We hope our approach could contribute to a deep understanding of self-assembly mechanisms and motivate new research on advanced multi-agent algorithms, massive collaboration mechanisms, and artificial collective intelligence systems.

\bibliographystyle{unsrt}

\section*{Acknowledgement}
Supported by the National Natural Science Foundation of China under grant numbers 61690200 and 61751210.

\newpage
\section*{Supplementary materials}
\textbf{The PDF file includes: } \\
Materials and Methods \\
Figs. S1 to S8 \\
Tables S1 to S4 \\
Captions for Movies S1 to S3 \\
Captions and Links for Dataset S1 \\
Captions and Links for Website S1 \\
\textbf{Other Supplementary Materials for this manuscript include the following: } \\
Movies S1 to S3 \\
Dataset S1 \\
Website S1 \\

\noindent
\textbf{Movie S1:} The self-assembly of shapes from 16 categories with thousands of agents in 80$\times$80 environment. One representative is selected for each shape category.

\noindent
\textbf{Movie S2:} The comparison of self-assembly processes of a randomly selected shape using different methods in 16$\times$16 environment with random initialization, 80$\times$80 environment with random initialization, and 80$\times$80 environment with specific initialization, respectively.

\noindent
\textbf{Movie S3:} The self-assembly of shapes with different scales. Each of the 4 representative shapes are formed in 6 environment scales with agents ranging from minimum 40 to maximum 5469.

\noindent
\textbf{Dataset S1:} The self-assembly shape set consisting of 156 shapes, each of which is represented by a black-white image of size 512$\times$512. Link: https://github.com/Catherine-Chu/Self-Assembly-Shape-Set.

\noindent
\textbf{Website S1:} The website for demonstrating self-assembly processes of the proposed approach. Link: http://self-assembly.qunzhi.fun.

\newpage

\setcounter{section}{0}
\section*{Materials and Methods}
\noindent
In the main text, we have demonstrated the performance of our approach for self-assembly with large-scale swarms. Here we provide more details about the problem formulation, the proposed algorithm, and the experiments.

Section 1 gives a formulation of the self-assembly problem. Section 2 presents in detail the proposed ALF-based self-assembly algorithm, including a formal definition of the ALF. Section 3 introduces more details of the experiments from 8 aspects: evaluation measures, baseline methods, the shape set used in experiments, experiment designs, parameter settings, experiment platforms, experimental results and analysis, discussion about weaknesses of baseline methods, and analysis of the influences of different parameter values on the performance of our approach.

\section{Problem Formulation}
The self-assembly problem focused in this paper involves three components: a grid environment $G$, a target shape $S$, and a group of agents $A$. The grid environment $G$ is defined as a matrix $\{(i,j)|i\in[1,H],j\in[1,W]\}$\footnote{For simplicity, a 2D grid environment is given here; however, the method proposed in this paper can naturally apply to 3D grid environments.}, where $H$/$W$ represents the height/width of the environment, and $(i,j)$ denotes the gird at row $i$ and column $j$. The target shape $S$ is defined as a subset of $G$ that forms a connected graph through the neighbor relation between grids, and let $|S|=N$. Grids in $S$ are called \emph{target} grids, and other grids \emph{un-target} grids. The group of agents $A$ is defined as a set $\{a_n|n\in[1,N]\}$, where $a_n$ denotes the agent with identity $n$. At any time, each agent occupies a distinct grid in $G$.

Agents interact with the environment in a sequence of discrete times: $0,1,...,t,...,T$. At each time $t$, each agent decides to stay at the current grid or move to one of its eight neighbor grids. When an agent decides to move and no conflict occurs, then in the next time $t+1$ the agent will appear at the new position; otherwise, the agent’s position will not be changed. 
The state of the system at time $t$, denoted as $p_t$, is an injective function from $A$ to $G$, mapping each agent to its occupied grid. At any time $t$, the group of agents $A$ is partitioned into two subsets: $I_t=\{a_n|a_n\in A,p_t(a_n)\in S\}$, and $O_t=A-I_t$. That is, $I_t$ consists of agents in shape $S$, and $O_t$ agents out of $S$. 
Accordingly, the target shape $S$ is partitioned into two subsets: $C_t=\{p_t(a_n)|a_n\in I_t\}$, and $U_t=S-C_t$.
Grids in $C_t$ are called \emph{occupied} target grids, and grids in $U_t$ \emph{unoccupied}.

\begin{figure*}[h!]
    \centering
    \includegraphics[scale=0.8,trim={0cm 0cm 0cm 0cm},clip]{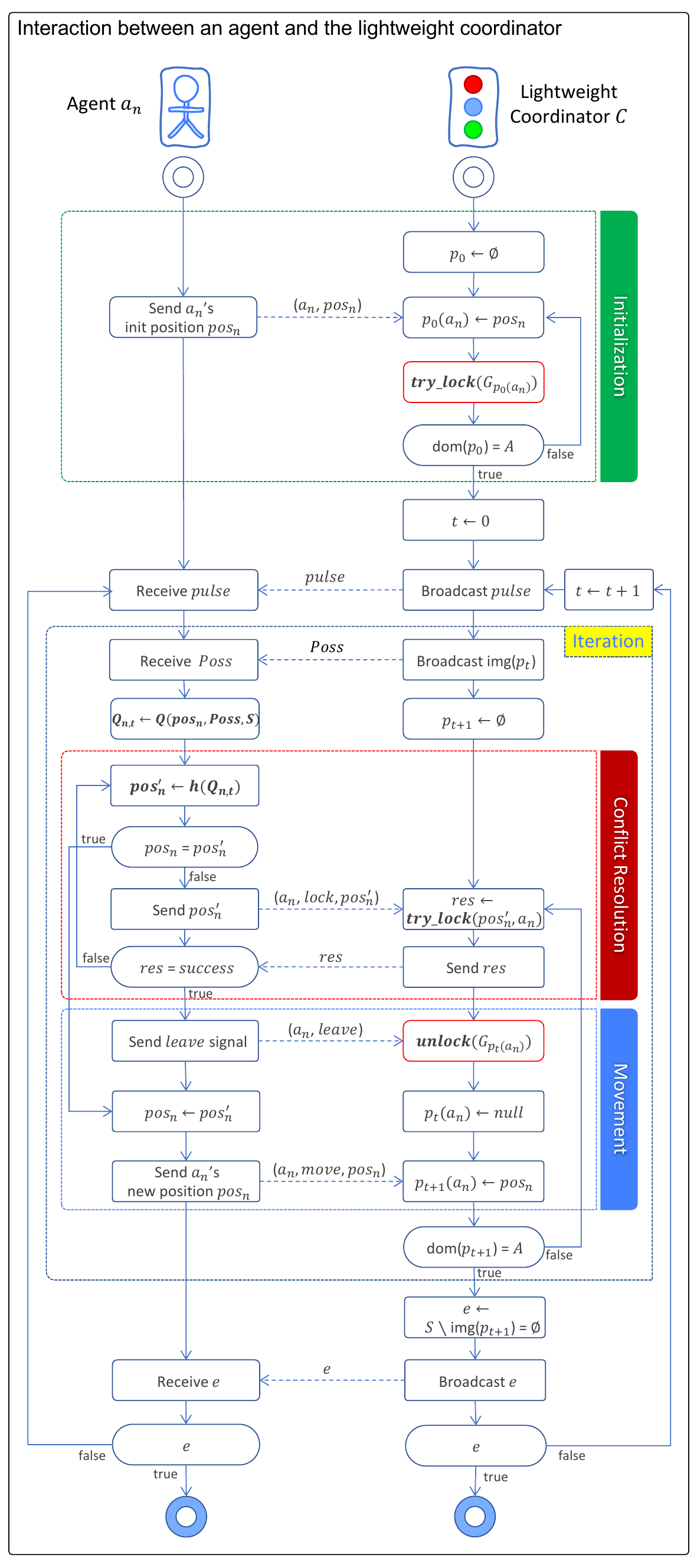}
    \begin{center}
        Figure S1: The interaction protocol between each agent and the lightweight coordinator.
    \end{center}
    \label{fig:interaction}
\end{figure*}

The goal of resolving this problem is to find a way from an initial state to a \emph{target} state as quickly as possible, following the interaction rule described above. In an initial state at time 0, all agents are randomly distributed in the environment. A target state is a state in which every target grid is occupied by an agent, i.e., $\mathit{img}(p_t)=S$.

\section{An ALF-based Self-Assembly Algorithm}
\subsection{Overview}
As mentioned in the main text, we design an artificial self-assembly system, consisting of five components: $G$, $S$, $A$, an \emph{artificial light field} $\mathcal{F}$ superimposed on $G$, and a \emph{lightweight coordinator} $C$. When resolving a self-assembly task, each agent interacts with $C$ through an iterative process (Figure S1), which consists of an initialization stage and a sequence of iteration stages corresponding to the sequence of system times. In the \emph{initialization} stage, 
\begin{itemize}
    \item[1)] each agent reports its initial position to $C$; (Algorithm 2 line 1-2)
    \item[2)] $C$ gets the system state at time 0, denoted as $p_0$. (Algorithm 3 line 1-2) 
\end{itemize}
In each \emph{iteration} at time $t$, 
\begin{itemize}
    \item[3)] $C$ broadcasts $\mathit{img}(p_t)$, i.e., all the positions occupied by agents, to each agent in $A$; (line 5 in Algorithm 3 and Algorithm 2)
    \item[4)] each agent calculates a priority queue of next positions (encapsulated in the $Q$ function); (line 6 in Algorithm 2)
    \item[5)] each agent sequentially retrieves elements from its queue and request $C$ to lock corresponding position for it, until finding a conflict-free next position (encapsulated in the $h$ function); (line 7-12 in Algorithm 2 and line 7-9 in Algorithm 3)
    \item[6)] the agent sends a leave signal before moving, updates its position, and then reports its new position to $C$, so that $C$ can update the system state accordingly; (line 13-15 in Algorithm 2 and line 10-14 in Algorithm 3)
    \item[7)] as the last step in each iteration, $C$ checks whether a target state is achieved, and triggers a new iteration at time $t+1$ if not or broadcasts an exit signal if true. (line 15-16 in Algorithm 3 and line 16, 4 in Algorithm 2)
\end{itemize}
\begin{algorithm*}[h]
    \setcounter{algocf}{0}
    \caption{System initialization}
    \label{alg:alf}
    \KwIn{$G$: a grid environment, $S$: a target shape, $A$: a group of agents, $p_0$: the system's initial state, $C$: a coordinator, $\gamma$: exploration rate, $\mathit{flag}$: whether agents can leave shape after entering it,
    $\mathcal{W}$: policy transformation parameter for agents in shape;
    } 
    Thread($C$).start($G$, $S$, $A$);\\
    \textbf{for} each $a_n \in A$ \textbf{do} Thread($a_n$).start($G$, $S$, $C$, $p_0(a_n)$, $\gamma$, $\mathit{flag}$, $\mathcal{W}$);
\end{algorithm*}
\begin{algorithm*}[htb]
    \setcounter{algocf}{1}
    \caption{Behavior of an agent $a_n$}  
    \label{alg:alf-a}  
    \KwIn{$G$: a grid environment, $S$: a target shape, $C$: a coordinator, $pos$: $a_n$'s initial position, $\gamma$: exploration rate, $\mathit{flag}$: whether agents can leave shape after entering it,
    $\mathcal{W}$: policy transformation parameter for agents in shape;
    }  
    sendInitPos($C$, $pos$);\\
    \textbf{let} $t \leftarrow 0$, \emph{pulse};\\
    \While{true}{
        $pulse \leftarrow$ recvPulse($C$); \textbf{if} \emph{pulse} = \emph{STOP} \textbf{then} break; \\
        \textbf{let} \emph{Poss} $\leftarrow$ recvPoss($C$);\\
        \textbf{let} \emph{Q} $\leftarrow$ calcPrefPosQueue($pos$, $S$, \emph{Poss}, $\mathit{flag}$, $\mathcal{W}$);\\
        \textbf{let} \emph{res} $\leftarrow$ \emph{FAIL}, \emph{prefPos}; \\ 
        \While{Q.\emph{empty()} = false \emph{and} res = FAIL}{
            \emph{prefPos} $\leftarrow$ $Q$.pop(); \\
            \textbf{if} \emph{prefPos} = $pos$ \textbf{then} rnd(0,1) $<\gamma$ ?  continue : break;\\
            sendPrefPosReq($C$, \emph{prefPos}); \\
            $res \leftarrow$ recvPrefPosRes($C$); \\
        }
        \If{res = SUCC}{
            sendLeaveSig($C$); $pos$ $\leftarrow$ \emph{prefPos}; \\
        }
        sendNewPos($C$, \emph{pos});\\
        $t$ $\leftarrow$ $t+1$; \\
    }
\end{algorithm*} 
\begin{algorithm*}[h]
    \setcounter{algocf}{2}
    \caption{Behavior of the coordinator}
    \label{alg:alf-c}
    \KwIn{$G$: a grid environment, $S$: a target shape, $A$: a group of agents;\\
    }
    \textbf{let} $p_0 \leftarrow$ recvPoss($A$); lockAll($p_0$);\\
    \textbf{let} $t$ $\leftarrow$ 0, $pulse \leftarrow WORK$;\\
    \While{true}{
        broadcastPulse($A$, \emph{pulse}); \textbf{if} \emph{pulse} = \emph{STOP} \textbf{then} break; \\
        broadcastPoss($A$, img($p_t$));\\
        \While{true}{
          \textbf{let} ($a_n$, \emph{msg}) $\leftarrow$ recvMsg($A$);\\
          \If{msg.type = PREF\_POS\_REQ}{
            sendPrefPosRes($a_n$, tryLock($a_n$, \emph{msg.value})); \\
          }\ElseIf{msg.type = LEAVE\_SIG}{
            unlock($p_t$($a_n$)); \\
          }\ElseIf{msg.type = NEW\_POS}{
            $p_{t+1}$($a_n$) $\leftarrow$ \emph{msg.value}; \\
            \textbf{if} all $a_n \in A$ finished actions at \emph{t} \textbf{then} break; \\
          }
        }

        \emph{pulse} $\leftarrow$ ($S \setminus \text{img}(p_{t+1}) = \emptyset$) ? \emph{STOP} : \emph{WORK}; \\
        $t \leftarrow t + 1$; \\
    }
\end{algorithm*}
	
\subsection{Artificial Light Field}
To support each agent calculating its priority queue of next positions, an artificial light field (ALF) is superimposed on the grid environment and updated dynamically according to the current system state. The ALF at time $t$, denoted as $\mathcal{F}_t$, is defined as a pair of functions $(r_t,b_t)$, where the former maps each grid to the intensity of red light at the grid, and the latter to the intensity of blue light at each grid.

The intensity of red/blue light at a grid is the sum of all red/blue light sources’ intensities at the grid. At any time t, each agent in $O_t$ is a source of red light, and each grid in $U_t$ is a source of blue light. The intensity of light from a source attenuates linearly with propagation distance. As a result, $r_t$ and $b_t$ can be defined conceptually as follows:
\begin{equation}
\begin{split}
    r_t\left(g\right)&=\sum_{a\in O_t}f\left(L,\alpha,dis\left(g,p_t\left(a\right)\right)\right),\ \ \ \ g\in G\\
    b_t\left(g\right)&=\sum_{g\prime\in U_t} f\left(L,\alpha,dis\left(g,g^\prime\right)\right),\ \ \ \ g\in G
\end{split}
\end{equation}
where $L$ is the intensity of light emitted by a light source, $\alpha$ is the attenuating rate of light, $\mathit{dis}$ is a function that returns the distance between two grids, and $f$ is a function that returns the intensity of light after the light has traveled a certain distance from its source.

At each time step, each agent will calculate its local light field based on the above equations, as presented in Algorithm 4.
\begin{algorithm*}[h]
\setcounter{algocf}{3}
    \caption{getNeiLightField}
    \label{alg:get_light}
    \KwIn{\\
    $\mathit{pos}$: the current position of $a_n$, 
    $S$: a target shape,
    $\mathit{Poss}$: the set of all agents' positions
    }
    \KwOut{\\
    $\mathcal{F}$: the blue and red light intensities at surrounding 8 grids and the current position;}
    \textbf{let} $\mathcal{F}$ = dict(), $L$, $\beta$; \\
    \textbf{let} $U_t = S \setminus Poss$, $O_t = Poss \setminus S$; \\
   
    \For{each $p$ in surrounding and current positions}{
        $b_{p}$ = $\sum_{g\in U_t} L/(1+ \beta \max_i(|p[i]-g[i]|))$; \\
        \If{pos $\in$ S}{
            $r_{p}$ = $\sum_{g\in O_t} L/(1+ \beta \max_i(|p[i]-g[i]|))$; \\
            $\mathcal{F}$[$p$] = ($b_{p}$, $r_{p}$); \\
        }\Else{
             $\mathcal{F}$[$p$] = $b_{p}$; \\
        }
    }
    \Return $\mathcal{F}$;
\end{algorithm*}

\subsection{The \textbf{\emph{Q}} Function for Generating Priority Queues}
The $Q$ function (defined in Algorithm 5) encapsulates an agent’s behavior strategy by returning the agent’s priority queue of next positions based on the agent’s local light field. Each element in the priority queue is either one of the agent’s eight neighbor positions or the agent’s current position, and no duplicate elements exists in the priority queue.

Two behavior strategies are designed for two kinds of agent state, respectively. When an agent $a_n\in O_t$, its priority queue of next positions is constructed by sorting all candidate positions in descending order of their intensities of blue light. The strategy drives an agent to move eagerly towards unoccupied target grids, as long as no conflicts occurs.
\begin{algorithm*}[h]
    \setcounter{algocf}{4}
    \caption{calcPrefPosQueue}  
    \label{alg:prior_queue}  
    \KwIn{\\
    $pos$: the current position of $a_n$,
    $S$: a target shape,
    $Poss$: the set of all agents' positions,
    $flag$: whether agents can leave shape after entering it,
    $\mathcal{W}$: policy transformation parameter for agents in shape;
    }  
    \KwOut{\\
    $Q$: priority queue of next positions;
    }
    \textbf{let} $W$ = $|Poss \setminus S|/|S|$, $Q$;\\
    \textbf{let} $\mathcal{F}$ = getNeiLightField($pos$, $S$, $Poss$); \\
    \If{$pos \in S$}{
        \If{$W > \mathcal{W}$}{
            $Q$.comp = $\textbf{bool}$ func($p_1,p_2$)
            \{\Return $\mathcal{F}[p_1].b<\mathcal{F}[p_2].b \, \,\mathbf{or}\,\,\mathcal{F}[p_1].b == \mathcal{F}[p_2].b\,\,\mathbf{and}\,\, \mathcal{F}[p_1].r>\mathcal{F}[p_2].r$;\}\;
        }\Else{
            $Q$.comp = $\textbf{bool}$ func($p_1$,$p_2$)
            \{\Return $\mathcal{F}[p_1].r>\mathcal{F}[p_2].r$;\}\;
        }
    }\Else{
        $Q$.comp = $\textbf{bool}$ func($p_1,p_2$)
        \{\Return $\mathcal{F}[p_1].b<\mathcal{F}[p_2].b$;\}\;
    }
    $Q$.push($pos$); \\ 
    \For{each p in surrounding 8 positions}{
        \If{$(pos \notin S) \,\,\mathbf{or}\,\, (pos \in S \,\,\mathbf{and}\,\, (p \in S \,\,\mathbf{or}\,\, (p \notin S \,\,\mathbf{and}\,\, \mathbf{not} \,\, flag)) )$}{
            $Q$.push(p);
        }
    }
    \Return $Q$ ;\\
\end{algorithm*} 

When an agent $a_n\in I_t$, its priority queue of next position will be constructed by two construction principles according to the task progress. The first principle obtains a priority queue by sorting all candidate positions in descending order of blue light intensity primarily, and in ascending order of red light intensity secondarily (line 4-5, 10-13 in Algorithm 5). This principle motivates an agent to keep moving towards the vacant position in the center of the shape, accelerating convergence at the beginning of the task. The second principle obtains a priority queue by sorting all candidate positions in ascending order of red light intensity (line 6-7, 10-13 in Algorithm 5). This strategy motivates an agent to leave the peripheral position open until convergence. For each agent in $I_t$, it uses the \emph{completion rate} $W\in\left[0,1\right]$ to decide which principle should be taken, where $W$ is defined as the proportion of occupied target grids to all target grids. When $W$ is less than a threshold $\mathcal{W}$, the agent will adopt the first construction principle; otherwise, the second principle will be adopted. In our experiments, we set the threshold  $\mathcal{W}=15\%$.

In Figure 1.D, for instance, $a_i$ is out of the shape and on the edge of the environment, so $a_i$ only has four candidate positions for the next step (including the current position of $a_i$, denoted as $p_t(a_i)$). In this case, since $a_i\in O_t$,  a priority queue is generated according to the first behavior strategy. The position below $p_t(a_i)$ has the highest blue light intensity, so it priors to all the other candidate positions. Specifically, although $p_t(a_i)$ has the same blue light intensity with the position at the right side of $p_t(a_i)$, we prior the position at the right side in the queue, because we always prefer agents to move. For another instance, $a_j$ is inside the shape and $W=6/11<\mathcal{W}$, so its priority queue is generated by the first construction principle of the second strategy. Regrading both blue and red light, the current position is the best choice.

\subsection{The \textbf{\emph{h}} and \textbf{\emph{try\_lock}} Functions for Conflict Avoidance}
The $h$ and \emph{try\_lock} functions encapsulate a decentralized strategy to mediate between different agents’ behavior by selecting an element from each agent’s priority queue as the agent’s next position, so as to avoid two kinds of conflict: (1) an agent moves to a next position that has been occupied by another agent; (2) two agents move to the same unoccupied next position. 

For each agent, it uses $h$ function to repeatedly retrieves elements from $Q_{n,t}$ , and sends request to $C$ for locking the corresponding position $\mathit{pos}$ for the agent until receiving a \emph{success} response. For the lightweight coordinator $C$, a lock mechanism is designed to achieve the goal of conflict avoidance. Each grid in $G$ is treated as an exclusive resource, and its accessibility is managed by a mutex lock. In the initialization stage of our approach, when receiving a position $p_0(a_n)$, $C$ locks the grid at $p_0(a_n)$ for $a_n$. In each iteration at time $t$, when receiving a leave signal from agent $a_n$, $C$ unlocks the grid at $p_t(a_n)$; when receiving a request from $a_n$ for locking next position $\mathit{pos}$, $C$ tries to lock the position for $a_n$ and then returns a success/fail response $\mathit{res}$ to $a_n$. 

Specially, when the retrieved element from $Q_{n,t}$ is the position of $p_t(a_n)$, since this position has been locked by the agent, the $h$ function has a $1-\gamma$ chance to directly return $p_t(a_n)$ as $a_n$’s next position, and a $\gamma$ chance to ignore $p_t(a_n)$ and continue retrieving the remaining elements after $p_t(a_n)$ (This stochastic strategy helps each agent to escape the local extremum). If all elements in $Q_{n,t}$ (except for $p_t(a_n))$ are inaccessible, the $h$ function will simply return $p_t(a_n)$. 

Consequently, after obtaining $p_{t+1}(a_n)$, agent $a_n$ will move to $p_{t+1}(a_n)$ and send a leave signal to $C$, causing $C$ to release the lock of $p_t(a_n)$. The \emph{try\_lock} mechanism is adopted in implementation to avoid dead lock caused by the simple \emph{lock\_until\_acquire} mechanism: when the \emph{try\_lock} is applied on an inaccessible grid, the locking process will immediately return a \emph{fail} result, so that the remaining positions in the priority queue can be checked timely.

\section{Self-Assembly Experiments}
\subsection{Evaluation Measures}
In the main text, we introduce three measures (\emph{completion quality} $\rho$ , \emph{relative completion time} $t$, and \emph{absolute completion time} $\tau$) to evaluate the performance of a self-assembly algorithm. The three measures in forming a shape are estimated through a set of repeated experiments, using the following equations:

Given a target shape $S$, a self-assembly algorithm $\mathcal{A}$, and $N$ repeated experiments for $\mathcal{A}$ to form $S$,
\begin{itemize}
    \item [1.] 	the completion quality is estimated by $\hat{\rho}\left(S,\mathcal{A},N\right)=\frac{1}{N}\sum_{e=1}^{N}\frac{|C_{S,\mathcal{A},e}|}{|S|}$, where $C_{S,\mathcal{A},e}$ denotes the set of occupied target grids when the $e$’th experiment terminates (the experiment terminates when either $S$ is formed, or the number of iterations exceeds a pre-defined threshold $K$);
    \item[2.] the relative completion time is estimated by $\hat{t}\left(S,\mathcal{A},N\right)=\frac{1}{\sum_{e=1}^{N}{\mathbf{1}(e)}}\sum_{e=1}^{N}{\mathbf{1}(e)T_{S,\mathcal{A},e}}$, where $T_{S,\mathcal{A},e}$ denotes the number of iterations to form shape $S$ by algorithm $\mathcal{A}$ in the $e$’th experiment; $\mathbf{1}(e)=1$ if $S$ is formed when the $e$’th experiment terminates, and 0 otherwise;
    \item[3.] 	the absolute completion time is estimated by $\hat{\tau}\left(S,\mathcal{A},N\right)=\frac{1}{\sum_{e=1}^{N}{\mathbf{1}(e)}}\sum_{e=1}^{N}{\mathbf{1}(e)\Gamma}_{S,\mathcal{A},e}$, where $\Gamma_{S,\mathcal{A},e}$ denotes the physical time to form shape $S$ by algorithm $\mathcal{A}$ in the $e$’th experiment.
\end{itemize}

In addition, the three measures in forming a group of shapes with different scales are estimated through multiple sets of repeated experiments, using the following equations:

Given a set of target shape $\mathcal{S}=\{S_1,...,S_M\}$, a self-assembly algorithm $\mathcal{A}$, and N repeated experiments for $\mathcal{A}$ to form $S_m$ $(m=1,2,..,M)$,
\begin{itemize}
    \item[1.] 	the completion quality is estimated by $\hat{\rho}\left(\mathcal{S},\mathcal{A},N\right)=\frac{1}{M}\sum_{m=1}^{M}{\hat{\rho}\left(S_m,\mathcal{A},N\right)}$;
    \item[2.] 	the relative completion time is estimated by  $\hat{t}\left(\mathcal{S},\mathcal{A},N\right)=\frac{1}{M}\sum_{m=1}^{M}\frac{|max(\mathcal{S})|}{|S_m|}\hat{t}\left(S_m,\mathcal{A},N\right)$, where $max(\mathcal{S})$ denotes the shape whose number of target grids is maximum in $\mathcal{S}$;
    \item[3.] 	the relative completion time is estimated by  $\hat{\tau}\left(\mathcal{S},\mathcal{A},N\right)=\frac{1}{M}\sum_{m=1}^{M}{\frac{|max(\mathcal{S})|}{|S_m|}\hat{\tau}\left(S_m,\mathcal{A},N\right)}$;
\end{itemize}

\subsection{Compared Methods}
To verify the advantage of our approach, we compare our approach with four state-of-the-art methods:

\textbf{Centralized Distance-Optimal Method (OPT-D)} \cite{JYu13}: this method uses a three-step process to resolve self-assembly problems: 1) calculates the shortest path between any pair of agent and target grid; 2) calculates a distance-optimal agent-grid assignment by Hungarian algorithm; 3) orders vertexes along paths and resolves conflicts by swapping the assigned girds of two conflicting agents. A significant property of OPT-D is that the maximum iteration to form a shape can be theoretically guaranteed to be $\left|A\right|+d_{\mathit{max}}-1$, where $|A|$ is the number of agents, and $d_{\mathit{max}}$ is the maximal minimal distance between agents and girds in a distance-optimal agent-grid assignment. One of the drawbacks of OPT-D is that the computational cost of global agent-gird assignment and vertex ordering will increase in $n^3$-form as the shape scale $n$ increases, causing poor scalability.

\textbf{Hungarian-Based Path Replanning (HUN)} \cite{alonso11}: this method uses an iterative process to resolve self-assembly problems, a process consisting of two alternating steps: 1) uses Hungarian algorithm to calculate a distance-optimal agent-grid assignment; 2) performs optimal reciprocal collision avoidance (ORCA) \cite{hcheng} to avoid local conflicts, until no agent can move without conflicts. These two steps are repeated until the shape is formed. In the case of sparse target grid distribution, this method can obtain near-optimal travel distances, whereas in the case of a dense target grid distribution, more iterations will be used for replanning. The cost of iterative replanning of targets and paths is high, causing both poor efficiency and scalability.  

\textbf{Dynamic Uniform Distribution (DUD)} \cite{bi18}: This method is based on the concept of artificial potential field (APF) \cite{chiang15}, which generally consists of an attraction field and a repulse field. In DUD, the distance-based gradients are used to generate the attraction field, and for any agent, the repulse field are only triggered when some other agent moves near the agent (i.e., the distance between the two agents is less than a pre-defined distance). The combined forces of attraction and repulsion directs agents to move until convergence. DUD has the advantage of high scalability, but suffers from the problem of local minima, causing poor stability. In addition, in our experiments, in order to apply the control strategy of DUD (which is originally designed for continuous environments) to discrete grid environments, two or more agents are allowed to occupy a same grid at the same time.

\textbf{Gradient-Based Edge-Following (E-F)} \cite{rubenstein14}: This method models each agent’s motion in self-assembly as an iterative edge-following process based on the gradient information towards seed grids. In initialization, four pre-localized seed grids of the target shape are settled and all agents are connected to the seed grids (directly or indirectly through other agents); in each iteration, two steps are carried out for each agent: 1.  the agent calculates its relative gradients towards seed grids; 2. if the agent finds it is at the outer edge (i.e., has a highest gradient value among all its neighbors), it will move along the edge clockwise (namely, \emph{edge-following}); otherwise, it will keep stationary. Each agent will continue its edge-following behavior until one of the two stop conditions is satisfied: 1. the agent has entered the target shape but is about to move out of the shape; 2. the agent is next to a stopped agent with the same or greater gradient. E-F has good scalability, but its efficiency and parallelism are limited due to the edge-following strategy.

\subsection{Shape Set}
To evaluate the performance of different methods on the self-assembly problem, we build a shape set\footnote{The shape set can be found at https://github.com/Catherine-Chu/Self-Assembly-Shape-Set.} with sufficient diversity. In particular, different sets are selected based on a shape classification as shown in the top of Figure S2:
\begin{itemize}
    \item[1.] The \emph{shape set} (containing 156 connected shapes) is divided into two subsets: \emph{shapes without holes}, and \emph{shapes with holes}, according to whether the shape has holes or not;
    \item[2.] The set of \emph{shapes without holes} (containing 87 shapes) is divided into two subsets: \emph{convex shapes} (containing 20 shapes), and \emph{concave shapes} (containing 67 shapes). The set of \emph{shapes with holes} (containing 69 shapes) is divided into two subsets: \emph{one-hole shapes} (containing 33 shapes) and \emph{multi-hole shapes} (containing 36 shapes), according to the number of holes;
    \item[3.] For each of the two sets of \emph{convex shapes} and \emph{concave shapes}, it is divided into three subsets: shapes enclosed by line segments (convex/concave \emph{line-enclosed}, containing 11/28 shapes), shapes enclosed by curves (convex/concave \emph{curve-enclosed}, containing 4/15 shapes), and shapes enclosed by both line segments $\&$ curves (convex/concave \emph{line $\&$ curve-enclosed}, containing 5/24 shapes), according to the smoothness of shape edge. The set of \emph{one-hole shapes} is divided into two sets: \emph{convex hole} (containing 21 shapes) and \emph{concave hole} (containing 12 shapes). The set of \emph{multi-hole shapes} is are divided into three subsets: \emph{convex holes} (containing 13 shapes), \emph{concave holes} (containing 10 shapes), and \emph{convex$\&$concave holes} (containing13 shapes), by composing multiple holes’ convexity and concavity;
    \item[4.] For each of the five sets of \emph{convex hole}, \emph{concave hole}, \emph{convex holes}, \emph{concave holes}, and \emph{convex$\&$concave holes}, it is divided into two subsets: \emph{convex contour} (containing 7/6/6/5/5 shapes), and \emph{concave contour} (containing 14/6/7/5/8 shapes), by the convexity and concavity of a shape’s contour.
\end{itemize}

\begin{figure}[h!]
    \centering
    \includegraphics[height=\dimexpr\pagegoal-\pagetotal\relax,width=\textwidth,keepaspectratio]{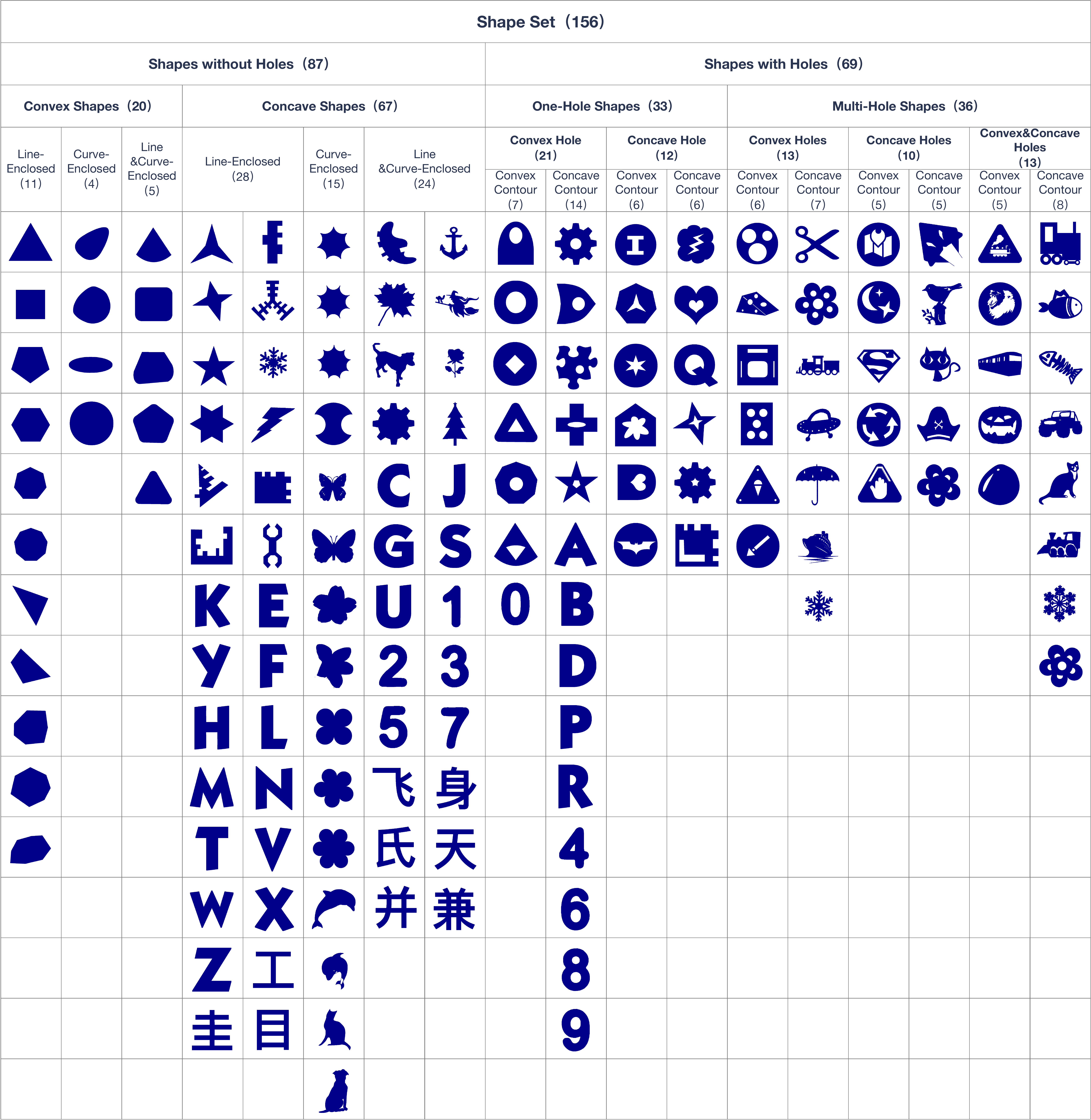}
    \begin{center}
        Figure S2. A shape classification and a shape set with 156 shapes.
    \end{center}
    \label{fig:shape-set}
    \end{figure}

In the shape set, each shape is represented as a 512$\times$512 black-white figure. A black pixel in a figure corresponds to a target grid, and a white pixel an un-target grid. In experiments, we zoom the figure into different scales on demand, from 15$\times$15 to 180$\times$180.

\subsection{Experiment Design}
To evaluate the efficiency of our approach, we compare it with the four baseline methods under two different policies of agents’ initial distribution: random and specific policies.
Experiments with the \emph{random} policy are carried out in three environments with scale of 16$\times$16, 40$\times$40, and 80$\times$80, respectively. In each environment, for each of the 156 shapes, we conduct 50, 50, 20, and 10 experiments for ALF, DUD, HUN, and OPT-D, respectively, and calculate three factors of completion quality $\rho$, relative completion time t, and absolute completion time $\tau$ of different methods. Experiments with the \emph{specific} policy are carried out in the  80$\times$80 environment for the 16 representative shapes listed in Table S1. For each of the 16 shapes, we conduct 50, 50, 20, 10 and 5 experiments for ALF, DUD, HUN, OPT-D and E-F, respectively, and the same measures with random-policy experiments are calculated.

To evaluate the scalability of our approach, we compare it with OPT-D on both relative completion time $t$ and absolute completion time $\tau$ for the 16 shapes (listed in Table.S1) in 12 different environment scales (the number of target grids varies from 40 to 5469, and the scale of environments varies from 15$\times$15 to 180$\times$180.). In particular, our approach is executed on two different hardware settings: a single-thread setting and a 16-thread setting. OPT-D is executed only on a single-thread setting, because it is not easy to transform OPT-D into a corresponding multi-thread version. We calculate $t$ and $\tau$ in each experiment, and analyze the changing trend of $t$ and $\tau$ as the number of target grids increases using two kinds of fitting: a \emph{linear} fitting and a \emph{log} fitting.

\begin{table}[h!]
\small
\centering
\begin{tabular}{|c|c|c|l|}
\hline
\textbf{\begin{tabular}[c]{@{}l@{}}Shape \\ ID (m)\end{tabular}} & \textbf{\begin{tabular}[c]{@{}l@{}}Shape\\  name\end{tabular}} & \textbf{\begin{tabular}[c]{@{}c@{}}Category\\ ID (k)\end{tabular}} & \textbf{\begin{tabular}[c]{@{}l@{}}Category Name\end{tabular}}                     \\ \hline
1                                                                      & 5-angles                                                             & 1                                                                         & Concave:line                               \\ \hline
2                                                                      & 4-curves                                                             & 2                                                                         & Concave:curve                              \\ \hline
3                                                                      & face                                                                 & 3                                                                         & Concave:line\&curve                         \\ \hline
4                                                                      & r-6-edge                                                             & 4                                                                         & Convex:line                                \\ \hline
5                                                                      & irre-curve-1                                                         & 5                                                                         & Convex:curve                               \\ \hline
6                                                                      & r-edge-3                                                             & 6                                                                         & Convex:line\&curve                          \\ \hline
7                                                                      & gear                                                                 & 7                                                                         & Hole:$<$Out$>$Concave$<$In$>$Convex                \\ \hline
8                                                                      & cloud\_lightning                                                      & 8                                                                         & Hole:$<$Out$>$Concave$<$In$>$Concave               \\ \hline
9                                                                      & end\_oval                                                             & 9                                                                         & Hole:$<$Out$>$Convex$<$In$>$Convex                 \\ \hline
10                                                                     & gong-bank                                                            & 10                                                                        & Hole:$<$Out$>$Convex$<$In$>$Concave                \\ \hline
11                                                                     & scissor                                                              & 11                                                                        & Multi-holes:$<$Out$>$Concave$<$In$>$Convex         \\ \hline
12                                                                     & aircraft                                                             & 12                                                                        & Multi-holes:$<$Out$>$Concave$<$In$>$Concave        \\ \hline
13                                                                     & locomotive                                                           & 13                                                                        & Multi-holes:$<$Out$>$Concave$<$In$>$Convex\&Concave \\ \hline
14                                                                     & maplog                                                               & 14                                                                        & Multi-holes:$<$Out$>$Convex$<$In$>$Concave         \\ \hline
15                                                                     & 3-holes                                                              & 15                                                                        & Multi-holes:$<$Out$>$Convex$<$In$>$Convex          \\ \hline
16                                                                     & train-roadsign                                                       & 16                                                                        & Multi-holes:$<$Out$>$Convex$<$In$>$Concave\&Convex  \\ \hline
\end{tabular}
\begin{center}\normalsize
    Table S1: The representative shapes from 16 categories in the shape set.
\end{center}
\end{table}

To evaluate the stability of our approach, we calculate the standard deviations of $\rho$, $t$, and $\tau$ of our approach on 50 randomly-initialized experiments for each of the 156 shapes in each of the three environments with scale of 16$\times$16, 40$\times$40, and 80$\times$80, respectively. In addition, for each shape in each environment, we also compare the relative completion time t of our approach with that of OPT-D.

\subsection{Parameter Settings}
\begin{equation}
    f\left(L,\beta,g,g^\prime\right)=
    \begin{cases}
    L-\beta \sum_{i=0}^{D-1}|g_i-g'_i| & \mathit{type}\,1 \\
    L-\beta \sqrt{\sum_{i=0}^{D-1}(g_i-g'_i)^2} & \mathit{type}\,2 \\
    L-\beta \max_{i\in[0,D)}|g_i-g'_i| & \mathit{type}\,3 \\
    \frac{L}{1+\beta \sum_{i=0}^{D-1}|g_i-g'_i|} & \mathit{type}\,4 \\
    \frac{L}{1+\beta \sqrt{\sum_{i=0}^{D-1}(g_i-g'_i)^2}} & \mathit{type}\,5 \\
    \frac{L}{1+\beta \max_{i\in[0,D)}|g_i-g'_i|} & \mathit{type}\,6 \\
    \frac{L}{1+\beta (\sum_{i=0}^{D-1}|g_i-g'_i|)^2} & \mathit{type}\,7 \\
    \frac{L}{1+\beta \sum_{i=0}{D-1}(g_i-g'_i)^2} & \mathit{type}\,8 \\
    \frac{L}{1+\beta (\max_{i\in[0,D)} |g_i-g'_i|)^2} & \mathit{type}\,9
    \end{cases}
\end{equation}

In all experiments in the main text, we set parameters $L=1000$, $\beta=1$, $\mathcal{W}=0.15$, $flag=True$, $\gamma=0.2$, where $L$ is the light intensity released by light sources, $\beta$ is the light discount coefficient, $\mathcal{W}$ is the threshold used to control the time of policy changing,  $\gamma$ is the agents’ exploration rate (when $\gamma=0$, it means the agent will never move to a position that is worse than the current one even there is no other better position to move), and flag denotes whether agents are allowed to move out of the shape after entering. For the light discount function $f$ (which describes the decay of light intensity with propagation distance), 9 different implementations are considered (see equation (2)), and the \textit{type} 6 implementation is used in our experiments. The 9 implementations of $f$ are generated by different combinations of two dimensions: the light-intensity distance-discount function, and the two-grid-distance measurement function. Three different values are considered in the former dimension: linear-discount (\textit{type} 1-3), inverse-discount (\textit{type} 4-6), and squared-inverse-discount (\textit{type} 7-9). Three different values are considered in the latter dimension: the Manhattan distance (\textit{type} 1, 4, 7), the European distance (\textit{type} 2, 5, 8), and the Chebyshev distance (\textit{type} 3, 6, 9).

The selection of these parameters is based on the observations of their effects on our algorithm’s performance through a set of experiments, in which 16 shapes listed in Table S1 are formed 50 times in 80$\times$80 environment with random initialization using different parameter-value combinations. In particular, we test 12$\times$6$\times$2$\times$9 different combinations of $\mathcal{W}$, $\gamma$, $\mathit{flag}$, and $f$ (12 values of  $\mathcal{W}$, 6 values of $\gamma$, 2 values of $\mathit{flag}$, and 9 values of $f$) for each of the 16 shapes. The combination of ($\mathcal{W}=0.15$, $\mathit{flag}=\mathit{True}$, $\gamma=0.2$, and $f=\mathit{type}$-6) achieves the minimum average relative completion time in the experiments, and we choose it as the default parameter values for our algorithm. The effects of each parameter on performance will be further discussed in Section 3.8.

\subsection{Experiment Platform}
All experiments are carried out on an HPC platform provided by Peking University, which can be accessed at http://hpc.pku.edu.cn/stat/wmyh. In particular, each experiment is carried out on a 16-cores HPC node (Intel Xeon E5-2697A V4) with 512G memory. 

\subsection{Results and Analysis}
\subsubsection{Efficiency}
The complete experimental results for evaluating efficiency are presented in Table S2-S4: 
\begin{itemize}
    \item[1.] Table S2 shows the shape-specific performance on $\hat{\rho}$, $\hat{t}$ and $\hat{\tau}$ of OPT-D, HUN, DUD and ALF (our approach) under random initialization policy for each of the 16 shapes (listed in Table S1) in 3 different environment scales. 
    \item[2.] Table S3 shows the category-specific performance on $\hat{\zeta}$,  $\hat{\rho}$, $\hat{t}$ and $\hat{\tau}$ of OPT-D, HUN, DUD and ALF under random initialization policy for each of the 16 categories (listed in Table S1), where $\hat{\zeta}$ denotes the success rate of all experiments in a category. 
    \item[3.] Table S4 shows the shape-specific performance on $\hat{\rho}$, $\hat{t}$ and $\hat{\tau}$ of OPT-D, HUN, DUD, E-F and ALF under specific initialization policy for each of the 16 shapes (listed in Table S1) in 80$\times$80 environment.
\end{itemize}
	
With random initialization policy, the results show that: 
\begin{itemize}
    \item[1.] For completion quality $\hat{\rho}$, OPT-D, HUN, DUD, and ALF (our approach) achieve 1.000, 0.913, 0.887, and 0.999 on average in all 156$\times$3 experiment settings, respectively, and the corresponding experiment success rate $\hat{\zeta}$ are 100\%, 26.90\%, 0\%, and 97.12\%, respectively.
    \item[2.] For relative completion time $\hat{t}$, OPT-D, HUN, and ALF achieve 211.28, 8653.21, and 640.64 iterations on average in all 156$\times$3 shapes, respectively; note that DUD fails in all 50 experiments for each shape in each environment, and HUN only successes on experiments in 16$\times$16 environments.
    \item[3.] For absolute completion time $\hat{\tau}$, OPT-D, HUN, and ALF achieve 1706.46, 70.66, and 12.24 seconds on average in all 156$\times$3 shapes, respectively; OPT-D spends most of the time in prior global task allocation and vertex ordering.
\end{itemize}

With specific initialization policy, the results show that: 
\begin{itemize}
    \item[1.] For completion quality $\hat{\rho}$, OPT-D, HUN, DUD, E-F, and ALF achieve 1.000, 0.907, 0.935, 0.739, and 0.999 on average in experiments for 16 shapes, respectively, and the corresponding experiment success rate $\hat{\zeta}$ are 100\%, 0\%, 0\%, 43.8\% and 87.5\%, respectively.
    \item[2.] For relative completion time $\hat{t}$, OPT-D, E-F, and ALF achieve 124.57, 7873.27, and 137.74 iterations on 16 shapes, respectively; HUN/DUD fails in all 20/50 experiments for each shape.
    \item[3.] For absolute completion time $\hat{\tau}$, OPT-D, E-F, and our approach achieve 1654.53, 732.71, and 29.56 seconds on 16 shapes, respectively. 
\end{itemize}
	
In summary, in terms of efficiency, our approach outperforms HUN, DUD, and E-F on all three measures; although performing worse than OPT-D on relative completion time, our approach achieves comparable completion quality and superior absolute completion time.

\clearpage
\newgeometry{left=1.4cm,right=1.4cm, top=2cm, bottom=2cm}
\begin{table}[h!]
\scriptsize
\begin{tabular}{|c|c|c|c|c|c|c|c|c|c|c|c|c|c|c|c|c|c|}
\hline
\multirow{2}{*}{\textbf{\begin{tabular}[c]{@{}c@{}}ID\\ (m)\end{tabular}}} & \multirow{2}{*}{\textbf{W/H}} & \multirow{2}{*}{\textbf{$\bm{|S_m|}$}} & \multicolumn{3}{c|}{\textbf{OPT-D}} & \multicolumn{3}{c|}{\textbf{HUN}} & \multicolumn{3}{c|}{\textbf{DUD}} & \multicolumn{6}{c|}{\textbf{ALF}}                                     \\ \cline{4-18} 
                                                                              &                               &                            & \textbf{$\bm{\hat\rho}$}  & \textbf{$\bm{\hat t}$}  & \textbf{$\bm{\hat\tau}$} & \textbf{$\bm{\hat\rho}$} & \textbf{$\bm{\hat t}$} & \textbf{$\bm{\hat\tau}$} & \textbf{$\bm{\hat\rho}$} & \textbf{$\bm{\hat t}$} & \textbf{$\bm{\hat\tau}$} & \textbf{$\bm{\hat\rho}$} & \textbf{$\bm{\hat t}$} & \textbf{$\bm{\hat\tau}$} & \textbf{$\bm{\sigma(\rho)}$} & \textbf{$\bm{\sigma(t)}$} & \textbf{$\bm{\sigma(\tau)}$} \\ \hline
\multirow{3}{*}{1}                                                            & 16                            & 31                         & 1.000      & 6.80       & 5.86      & 1.000     & 40.85     & 0.26      & 0.691     & -         & -         & 1.000     & 9.18      & 0.01      & 0.00000   & 0.05161   & 0.00001   \\ \cline{2-18} 
                                                                              & 40                            & 228                        & 1.000      & 17.80      & 120.79    & 0.875     & -         & -         & 0.732     & -         & -         & 1.000     & 21.66     & 0.34      & 0.00000   & 0.00204   & 0.00001   \\ \cline{2-18} 
                                                                              & 80                            & 930                        & 1.000      & 36.60      & 250.04    & 0.907     & -         & -         & 0.864     & -         & -         & 1.000     & 44.50     & 8.81      & 0.00000   & 0.01404   & 0.00091   \\ \hline
\multirow{3}{*}{2}                                                            & 16                            & 43                         & 1.000      & 7.60       & 6.12      & 1.000     & 47.20     & 0.12      & 0.738     & -         & -         & 1.000     & 11.45     & 0.02      & 0.00000   & 0.05116   & 0.00001   \\ \cline{2-18} 
                                                                              & 40                            & 315                        & 1.000      & 19.40      & 170.81    & 0.876     & -         & -         & 0.895     & -         & -         & 1.000     & 36.30     & 0.68      & 0.00000   & 0.02254   & 0.00011   \\ \cline{2-18} 
                                                                              & 80                            & 1316                       & 1.000      & 47.00      & 798.04    & 0.926     & -         & -         & 0.825     & -         & -         & 1.000     & 110.00    & 25.16     & 0.00000   & 0.01596   & 0.00048   \\ \hline
\multirow{3}{*}{3}                                                            & 16                            & 54                         & 1.000      & 8.00       & 5.55      & 0.980     & 373.55    & 1.56      & 0.890     & -         & -         & 1.000     & 11.60     & 0.02      & 0.00000   & 0.03704   & 0.00001   \\ \cline{2-18} 
                                                                              & 40                            & 322                        & 1.000      & 19.60      & 174.97    & 0.855     & -         & -         & 0.944     & -         & -         & 1.000     & 27.82     & 0.55      & 0.00000   & 0.01149   & 0.00007   \\ \cline{2-18} 
                                                                              & 80                            & 1296                       & 1.000      & 43.40      & 697.87    & 0.910     & -         & -         & 0.956     & -         & -         & 1.000     & 70.63     & 15.39     & 0.00000   & 0.00610   & 0.00079   \\ \hline
\multirow{3}{*}{4}                                                            & 16                            & 65                         & 1.000      & 6.80       & 5.32      & 0.837     & 488.00    & 3.82      & 0.917     & -         & -         & 1.000     & 10.18     & 0.04      & 0.00000   & 0.03231   & 0.00001   \\ \cline{2-18} 
                                                                              & 40                            & 404                        & 1.000      & 19.00      & 226.24    & 0.896     & -         & -         & 0.962     & -         & -         & 1.000     & 23.88     & 1.05      & 0.00000   & 0.00792   & 0.00005   \\ \cline{2-18} 
                                                                              & 80                            & 1592                       & 1.000      & 54.00      & 13761.19  & 0.939     & -         & -         & 0.978     & -         & -         & 1.000     & 48.13     & 2.68      & 0.00000   & 0.00327   & 0.00014   \\ \hline
\multirow{3}{*}{5}                                                            & 16                            & 47                         & 1.000      & 6.60       & 7.94      & 0.987     & 371.90    & 2.59      & 0.929     & -         & -         & 1.000     & 9.56      & 0.01      & 0.00000   & 0.05106   & 0.00001   \\ \cline{2-18} 
                                                                              & 40                            & 303                        & 1.000      & 17.00      & 162.99    & 0.879     & -         & -         & 0.962     & -         & -         & 1.000     & 23.20     & 0.43      & 0.00000   & 0.01023   & 0.00009   \\ \cline{2-18} 
                                                                              & 80                            & 1223                       & 1.000      & 44.40      & 565.39    & 0.926     & -         & -         & 0.980     & -         & -         & 1.000     & 48.17     & 11.06     & 0.00000   & 0.00392   & 0.00044   \\ \hline
\multirow{3}{*}{6}                                                            & 16                            & 41                         & 1.000      & 6.80       & 5.59      & 0.944     & 164.35    & 1.52      & 0.889     & -         & -         & 1.000     & 9.58      & 0.03      & 0.00000   & 0.04878   & 0.00002   \\ \cline{2-18} 
                                                                              & 40                            & 272                        & 1.000      & 21.40      & 146.16    & 0.886     & -         & -         & 0.947     & -         & -         & 1.000     & 22.68     & 0.87      & 0.00000   & 0.01029   & 0.00009   \\ \cline{2-18} 
                                                                              & 80                            & 1088                       & 1.000      & 41.60      & 416.07    & 0.920     & -         & -         & 0.973     & -         & -         & 1.000     & 50.42     & 2.17      & 0.00000   & 0.00607   & 0.00031   \\ \hline
\multirow{3}{*}{7}                                                            & 16                            & 60                         & 1.000      & 5.80       & 5.58      & 0.947     & 126.65    & 1.42      & 0.951     & -         & -         & 1.000     & 9.46      & 0.02      & 0.00000   & 0.02333   & 0.00001   \\ \cline{2-18} 
                                                                              & 40                            & 394                        & 1.000      & 16.40      & 218.38    & 0.876     & -         & -         & 0.969     & -         & -         & 1.000     & 22.96     & 0.55      & 0.00000   & 0.00711   & 0.00005   \\ \cline{2-18} 
                                                                              & 80                            & 1585                       & 1.000      & 38.60      & 1220.28   & 0.926     & -         & -         & 0.968     & -         & -         & 1.000     & 53.88     & 15.96     & 0.00000   & 0.00391   & 0.00021   \\ \hline
\multirow{3}{*}{8}                                                            & 16                            & 71                         & 1.000      & 6.80       & 5.76      & 0.835     & 399.00    & 4.29      & 0.904     & -         & -         & 1.000     & 9.48      & 0.02      & 0.00000   & 0.00423   & 0.00001   \\ \cline{2-18} 
                                                                              & 40                            & 434                        & 1.000      & 18.60      & 244.28    & 0.900     & -         & -         & 0.950     & -         & -         & 1.000     & 76.84     & 1.82      & 0.00000   & 0.00461   & 0.00030   \\ \cline{2-18} 
                                                                              & 80                            & 1729                       & 1.000      & 48.40      & 1738.00   & 0.927     & -         & -         & 0.973     & -         & -         & 0.997     & 89.57     & 23.36     & 0.00314   & 0.00607   & 0.00118   \\ \hline
\multirow{3}{*}{9}                                                            & 16                            & 75                         & 1.000      & 7.20       & 5.68      & 0.887     & 464.50    & 2.67      & 0.915     & -         & -         & 1.000     & 10.54     & 0.02      & 0.00000   & 0.06000   & 0.00001   \\ \cline{2-18} 
                                                                              & 40                            & 481                        & 1.000      & 22.40      & 280.15    & 0.903     & -         & -         & 0.952     & -         & -         & 1.000     & 26.94     & 0.73      & 0.00000   & 0.00644   & 0.00007   \\ \cline{2-18} 
                                                                              & 80                            & 1886                       & 1.000      & 57.80      & 2324.03   & 0.928     & -         & -         & 0.973     & -         & -         & 1.000     & 58.25     & 19.27     & 0.00000   & 0.00233   & 0.00018   \\ \hline
\multirow{3}{*}{10}                                                           & 16                            & 60                         & 1.000      & 5.40       & 260.43    & 1.000     & 31.60     & 0.12      & 0.723     & -         & -         & 1.000     & 13.35     & 0.02      & 0.00000   & 0.05000   & 0.00002   \\ \cline{2-18} 
                                                                              & 40                            & 430                        & 1.000      & 17.00      & 242.81    & 0.881     & -         & -         & 0.905     & -         & -         & 1.000     & 44.40     & 1.01      & 0.00000   & 0.01163   & 0.00020   \\ \cline{2-18} 
                                                                              & 80                            & 1722                       & 1.000      & 44.20      & 1466.37   & 0.915     & -         & -         & 0.913     & -         & -         & 1.000     & 180.00    & 47.03     & 0.00000   & 0.00720   & 0.00151   \\ \hline
\multirow{3}{*}{11}                                                           & 16                            & 37                         & 1.000      & 6.60       & 7.89      & 1.000     & 23.80     & 0.18      & 0.564     & -         & -         & 1.000     & 61.58     & 0.03      & 0.00000   & 0.05946   & 0.00001   \\ \cline{2-18} 
                                                                              & 40                            & 231                        & 1.000      & 17.20      & 122.29    & 0.827     & -         & -         & 0.925     & -         & -         & 1.000     & 10.06     & 0.91      & 0.00000   & 0.01645   & 0.00001   \\ \cline{2-18} 
                                                                              & 80                            & 907                        & 1.000      & 36.20      & 210.48    & 0.868     & -         & -         & 0.937     & -         & -         & 1.000     & 23.48     & 1.84      & 0.00000   & 0.00915   & 0.00010   \\ \hline
\multirow{3}{*}{12}                                                           & 16                            & 72                         & 1.000      & 9.20       & 5.69      & 0.900     & 519.80    & 5.01      & 0.849     & -         & -         & 1.000     & 11.64     & 0.02      & 0.00000   & 0.02500   & 0.00001   \\ \cline{2-18} 
                                                                              & 40                            & 458                        & 1.000      & 19.60      & 264.93    & 0.879     & -         & -         & 0.941     & -         & -         & 1.000     & 35.68     & 0.94      & 0.00000   & 0.01048   & 0.00008   \\ \cline{2-18} 
                                                                              & 80                            & 1865                       & 1.000      & 45.80      & 2002.64   & 0.923     & -         & -         & 0.903     & -         & -         & 1.000     & 89.21     & 29.16     & 0.00000   & 0.00901   & 0.00042   \\ \hline
\multirow{3}{*}{13}                                                           & 16                            & 42                         & 1.000      & 6.20       & 6.12      & 1.000     & 41.40     & 0.09      & 0.684     & -         & -         & 1.000     & 16.45     & 0.04      & 0.00000   & 0.02857   & 0.00001   \\ \cline{2-18} 
                                                                              & 40                            & 451                        & 1.000      & 21.20      & 248.42    & 0.919     & -         & -         & 0.873     & -         & -         & 1.000     & 29.40     & 1.84      & 0.00000   & 0.01086   & 0.00010   \\ \cline{2-18} 
                                                                              & 80                            & 1785                       & 1.000      & 51.40      & 1720.65   & 0.914     & -         & -         & 0.802     & -         & -         & 1.000     & 72.60     & 3.21      & 0.00000   & 0.00521   & 0.00106   \\ \hline
\multirow{3}{*}{14}                                                           & 16                            & 65                         & 1.000      & 6.40       & 10.18     & 1.000     & 83.00     & 0.40      & 0.803     & -         & -         & 1.000     & 21.24     & 0.08      & 0.00215   & 0.03231   & 0.00005   \\ \cline{2-18} 
                                                                              & 40                            & 425                        & 1.000      & 12.00      & 227.95    & 0.843     & -         & -         & 0.913     & -         & -         & 1.000     & 43.32     & 6.34      & 0.00000   & 0.00400   & 0.00014   \\ \cline{2-18} 
                                                                              & 80                            & 1655                       & 1.000      & 28.40      & 1236.88   & 0.888     & -         & -         & 0.954     & -         & -         & 0.999     & 245.50    & 3.01      & 0.00192   & 0.00393   & 0.00281   \\ \hline
\multirow{3}{*}{15}                                                           & 16                            & 58                         & 1.000      & 5.60       & 5.66      & 1.000     & 33.15     & 0.11      & 0.724     & -         & -         & 1.000     & 11.35     & 0.02      & 0.00000   & 0.02759   & 0.00001   \\ \cline{2-18} 
                                                                              & 40                            & 438                        & 1.000      & 17.80      & 247.54    & 0.850     & -         & -         & 0.906     & -         & -         & 1.000     & 25.95     & 0.69      & 0.00000   & 0.01164   & 0.00006   \\ \cline{2-18} 
                                                                              & 80                            & 1670                       & 1.000      & 38.60      & 1348.07   & 0.910     & -         & -         & 0.924     & -         & -         & 1.000     & 64.25     & 19.51     & 0.00000   & 0.00641   & 0.00053   \\ \hline
\multirow{3}{*}{16}                                                           & 16                            & 51                         & 1.000      & 6.00       & 5.53      & 1.000     & 54.35     & 0.59      & 0.910     & -         & -         & 1.000     & 9.02      & 0.04      & 0.00000   & 0.03333   & 0.00000   \\ \cline{2-18} 
                                                                              & 40                            & 346                        & 1.000      & 22.60      & 186.65    & 0.859     & -         & -         & 0.928     & -         & -         & 1.000     & 30.66     & 3.55      & 0.00000   & 0.01590   & 0.00009   \\ \cline{2-18} 
                                                                              & 80                            & 1418                       & 1.000      & 57.80      & 920.26    & 0.904     & -         & -         & 0.944     & -         & -         & 1.000     & 165.67    & 2.51      & 0.00000   & 0.01537   & 0.00073   \\ \hline
\end{tabular}
\begin{center}\normalsize
    Table S2: The shape-specific self-assembly performance of different methods with random initialization policy in three environment scales.
\end{center}
\end{table}
\restoregeometry

\clearpage
\begin{landscape}
\begin{table}[h!]
\scriptsize
\centering
\begin{tabular}{|c|c|c|c|c|c|c|c|c|c|c|c|c|c|c|c|c|c|c|c|c|}
\hline
\multirow{2}{*}{\textbf{\begin{tabular}[c]{@{}c@{}}ID\\ (k)\end{tabular}}} & \multirow{2}{*}{\textbf{$\bm{|S_k|}$}} & \multicolumn{4}{c|}{\textbf{OPT-D}}                                     & \multicolumn{4}{c|}{\textbf{HUN}}                                     & \multicolumn{4}{c|}{\textbf{DUD}}                       & \multicolumn{7}{c|}{\textbf{ALF}}                                                                                             \\ \cline{3-21} 
                             &                            & \textbf{$\bm{\hat \zeta}$}         & \textbf{$\bm{\hat\rho}$}      & \textbf{$\bm{\hat t}$}       & \textbf{$\bm{\hat \tau}$}        & \textbf{$\bm{\hat \zeta}$}        & \textbf{$\bm{\hat\rho}$}      & \textbf{$\bm{\hat t}$}        & \textbf{$\bm{\hat \tau}$}      & \textbf{$\bm{\hat \zeta}$}    & \textbf{$\bm{\hat\rho}$}      & \textbf{$\bm{\hat t}$}  & \textbf{$\bm{\hat \tau}$}  & \textbf{$\bm{\hat \zeta}$}        & \textbf{$\bm{\hat\rho}$}      & \textbf{$\bm{\hat t}$}       & \textbf{$\bm{\hat \tau}$}      & \textbf{$\bm{\sigma(\rho)}$} & \textbf{$\bm{\sigma(t)}$} & \textbf{$\bm{\sigma(\tau)}$}      \\ \hline
1                            & 28                         & 100.00\%          & 1.000          & 172.66          & 909.26           & 32.00\%          & 0.920          & 4559.84          & 43.35          & 0\%          & 0.866          & -          & -          & 99.63\%          & 1.000          & 439.91          & 7.93           & 0.00002          & 0.02891          & 0.00037          \\ \hline
2                            & 15                         & 100.00\%          & 1.000          & 184.02          & 834.74           & 22.96\%          & 0.911          & 7116.26          & 62.23          & 0\%          & 0.892          & -          & -          & 100.00\%         & 1.000          & 268.25          & 7.77           & 0.00000          & 0.01797          & 0.00023          \\ \hline
3                            & 24                         & 100.00\%          & 1.000          & 223.00          & 995.52           & 33.72\%          & 0.916          & 6404.34          & 43.48          & 0\%          & 0.849          & -          & -          & 93.15\%          & 0.999          & 527.28          & 7.08           & 0.00052          & 0.07800          & 0.00047          \\ \hline
4                            & 11                         & 100.00\%          & 1.000          & 139.62          & 1612.37          & 21.11\%          & 0.916          & 9431.11          & 69.18          & 0\%          & 0.938          & -          & -          & 100.00\%         & 1.000          & 280.94          & 4.82           & 0.00000          & 0.01452          & 0.00011          \\ \hline
5                            & 4                          & 100.00\%          & 1.000          & 197.40          & 1372.27          & 16.94\%          & 0.917          & 10520.29         & 80.68          & 0\%          & 0.948          & -          & -          & 100.00\%         & 1.000          & 376.44          & 9.16           & 0.00000          & 0.01749          & 0.00014          \\ \hline
6                            & 5                          & 100.00\%          & 1.000          & 145.25          & 1032.81          & 14.44\%          & 0.902          & 12765.33         & 99.28          & 0\%          & 0.948          & -          & -          & 100.00\%         & 1.000          & 380.51          & 3.72           & 0.00000          & 0.01327          & 0.00009          \\ \hline
7                            & 14                         & 100.00\%          & 1.000          & 140.55          & 2299.51          & 29.47\%          & 0.923          & 7233.13          & 63.77          & 0\%          & 0.910          & -          & -          & 100.00\%         & 1.000          & 361.94          & 7.37           & 0.00000          & 0.02031          & 0.00017          \\ \hline
8                            & 6                          & 100.00\%          & 1.000          & 150.68          & 4092.29          & 19.44\%          & 0.906          & 12197.25         & 123.90         & 0\%          & 0.906          & -          & -          & 83.35\%          & 0.997          & 865.19          & 12.21          & 0.00069          & 0.09087          & 0.00084          \\ \hline
9                            & 7                          & 100.00\%          & 1.000          & 121.85          & 2093.70          & 22.22\%          & 0.913          & 8920.09          & 83.82          & 0\%          & 0.939          & -          & -          & 99.80\%          & 1.000          & 464.18          & 7.06           & 0.00002          & 0.02406          & 0.00026          \\ \hline
10                           & 6                          & 100.00\%          & 1.000          & 111.17          & 2960.16          & 18.52\%          & 0.901          & 10907.08         & 100.71         & 0\%          & 0.922          & -          & -          & 92.63\%          & 0.999          & 278.49          & 23.85          & 0.00104          & 0.05642          & 0.00050          \\ \hline
11                           & 7                          & 100.00\%          & 1.000          & 189.37          & 580.19           & 35.08\%          & 0.918          & 3459.76          & 19.37          & 0\%          & 0.831          & -          & -          & 95.04\%          & 0.998          & 584.74          & 4.98           & 0.00037          & 0.03597          & 0.00031          \\ \hline
12                           & 5                          & 100.00\%          & 1.000          & 141.06          & 882.96           & 19.56\%          & 0.900          & 9173.47          & 78.49          & 0\%          & 0.869          & -          & -          & 92.78\%          & 0.999          & 310.46          & 24.86          & 0.00119          & 0.03085          & 0.00047          \\ \hline
13                           & 8                          & 100.00\%          & 1.000          & 157.02          & 1197.09          & 31.21\%          & 0.902          & 4784.38          & 31.34          & 0\%          & 0.819          & -          & -          & 92.56\%          & 0.999          & 358.47          & 7.91           & 0.00029          & 0.02375          & 0.00040          \\ \hline
14                           & 5                          & 100.00\%          & 1.000          & 133.96          & 981.44           & 26.00\%          & 0.906          & 6159.94          & 65.61          & 0\%          & 0.891          & -          & -          & 73.96\%          & 0.990          & 2234.82         & 16.64          & 0.00156          & 0.01717          & 0.00034          \\ \hline
15                           & 6                          & 100.00\%          & 1.000          & 141.79          & 1286.51          & 22.41\%          & 0.902          & 8551.07          & 76.11          & 0\%          & 0.907          & -          & -          & 100.00\%         & 1.000          & 230.17          & 15.35          & 0.00000          & 0.24837          & 0.00019          \\ \hline
16                           & 5                          & 100.00\%          & 1.000          & 129.99          & 1032.40          & 27.96\%          & 0.915          & 7080.13          & 53.52          & 0\%          & 0.912          & -          & -          & 89.67\%          & 0.998          & 1192.29         & 23.39          & 0.00203          & 0.10819          & 0.00030          \\ \hline
\textbf{All}                 & \textbf{156}               & \textbf{100.00\%} & \textbf{1.000} & \textbf{211.28} & \textbf{1706.46} & \textbf{26.90\%} & \textbf{0.913} & \textbf{8653.21} & \textbf{70.66} & \textbf{0\%} & \textbf{0.887} & \textbf{-} & \textbf{-} & \textbf{97.12\%} & \textbf{0.999} & \textbf{640.64} & \textbf{12.24} & \textbf{0.00034} & \textbf{0.04410} & \textbf{0.00033} \\ \hline
\end{tabular}
\begin{center}\normalsize
    Table S3: The category-specific self-assembly performance of different methods with random initialization policy.
\end{center}
\end{table}

\begin{table}[h!]
\scriptsize
\centering
\begin{tabular}{|c|c|c|c|c|c|c|c|c|c|c|c|c|c|c|c|c|}
\hline
\multirow{2}{*}{\textbf{\begin{tabular}[c]{@{}c@{}}ID\\ (m)\end{tabular}}} & \multirow{2}{*}{\textbf{$\bm{|S_m|}$}\footnotemark[3]} & \multicolumn{3}{c|}{\textbf{OPT-D}}                 & \multicolumn{3}{c|}{\textbf{HUN}}        & \multicolumn{3}{c|}{\textbf{DUD}}        & \multicolumn{3}{c|}{\textbf{E-F}}                   & \multicolumn{3}{c|}{\textbf{ALF}}                 \\ \cline{3-17} 
                                                                           &                            & \textbf{}      & \textbf{}       & \textbf{}        & \textbf{}      & \textbf{}  & \textbf{}  & \textbf{}      & \textbf{}  & \textbf{}  & \textbf{}      & \textbf{}        & \textbf{}       & \textbf{}      & \textbf{}       & \textbf{}      \\ \hline
1                                                                          & 930                        & 1.000          & 83.00           & 360.75           & 0.907          & -          & -          & 0.897          & -          & -          & 0.866          & 3628.20          & 293.39          & 1.000          & 67.46           & 9.63           \\ \hline
2                                                                          & 1316                       & 1.000          & 107.00          & 1023.95          & 0.912          & -          & -          & 0.966          & -          & -          & 0.951          & 6020.00          & 547.18          & 1.000          & 96.50           & 18.58          \\ \hline
3                                                                          & 1296                       & 1.000          & 82.00           & 887.78           & 0.900          & -          & -          & 0.937          & -          & -          & 0.938          & -                & -               & 1.000          & 94.98           & 18.83          \\ \hline
4                                                                          & 1592                       & 1.000          & 130.00          & 1644.93          & 0.914          & -          & -          & 0.944          & -          & -          & 0.930          & 6155.80          & 703.90          & 1.000          & 75.42           & 17.44          \\ \hline
5                                                                          & 1223                       & 1.000          & 47.00           & 696.21           & 0.910          & -          & -          & 0.963          & -          & -          & 0.062          & 4639.00          & 403.33          & 0.993          & 65.48           & 10.79          \\ \hline
6                                                                          & 1088                       & 1.000          & 56.00           & 520.31           & 0.925          & -          & -          & 0.960          & -          & -          & 0.895          & 4818.00          & 367.07          & 1.000          & 59.54           & 8.95           \\ \hline
7                                                                          & 1585                       & 1.000          & 124.00          & 1602.27          & 0.914          & -          & -          & 0.941          & -          & -          & 0.281          & -                & -               & 1.000          & 79.28           & 19.62          \\ \hline
8                                                                          & 1729                       & 1.000          & 86.00           & 1686.82          & 0.909          & -          & -          & 0.947          & -          & -          & 0.312          & -                & -               & 1.000          & -               & -              \\ \hline
9                                                                          & 1886                       & 1.000          & 58.00           & 2352.07          & 0.887          & -          & -          & 0.944          & -          & -          & 0.947          & -                & -               & 0.992          & 65.18           & 18.17          \\ \hline
10                                                                         & 1722                       & 1.000          & 66.00           & 1854.67          & 0.927          & -          & -          & 0.957          & -          & -          & 0.969          & 6876.60          & 1053.00         & 1.000          & -               & -              \\ \hline
11                                                                         & 907                        & 1.000          & 101.00          & 357.14           & 0.916          & -          & -          & 0.926          & -          & -          & 0.021          & 4231.00          & 242.56          & 1.000          & 118.66          & 18.70          \\ \hline
12                                                                         & 1865                       & 1.000          & 179.00          & 2542.45          & 0.883          & -          & -          & 0.866          & -          & -          & 0.896          & -                & -               & 1.000          & 148.34          & 42.50          \\ \hline
13                                                                         & 1785                       & 1.000          & 84.00           & 1895.45          & 0.932          & -          & -          & 0.900          & -          & -          & 0.973          & -                & -               & 1.000          & 121.74          & 32.80          \\ \hline
14                                                                         & 1655                       & 1.000          & 141.00          & 1773.43          & 0.920          & -          & -          & 0.934          & -          & -          & 0.958          & -                & -               & 1.000          & 232.66          & 59.02          \\ \hline
15                                                                         & 1670                       & 1.000          & 88.00           & 1766.94          & 0.870          & -          & -          & 0.955          & -          & -          & 0.892          & -                & -               & 1.000          & 75.48           & 19.45          \\ \hline
16                                                                         & 1418                       & 1.000          & 88.00           & 1198.82          & 0.883          & -          & -          & 0.915          & -          & -          & 0.932          & -                & -               & 1.000          & 140.02          & 30.29          \\ \hline
\textbf{All}                                                               & \textbf{-}                 & \textbf{1.000} & \textbf{124.57} & \textbf{1654.53} & \textbf{0.907} & \textbf{-} & \textbf{-} & \textbf{0.935} & \textbf{-} & \textbf{-} & \textbf{0.739} & \textbf{7873.27} & \textbf{732.71} & \textbf{0.999} & \textbf{137.74} & \textbf{29.56} \\ \hline
\end{tabular}
\begin{center}\normalsize
    Table S4: The shape-specific self-assembly performance of different methods with specific initialization policy in  80$\times$80 environments.
\end{center}
\end{table}
\end{landscape}
\clearpage

\restoregeometry
\normalsize
\baselineskip22pt

\subsubsection{Scalability}
\begin{figure}[h!]
\centering
\includegraphics[height=\dimexpr\pagegoal-\pagetotal\relax,width=\textwidth,keepaspectratio]{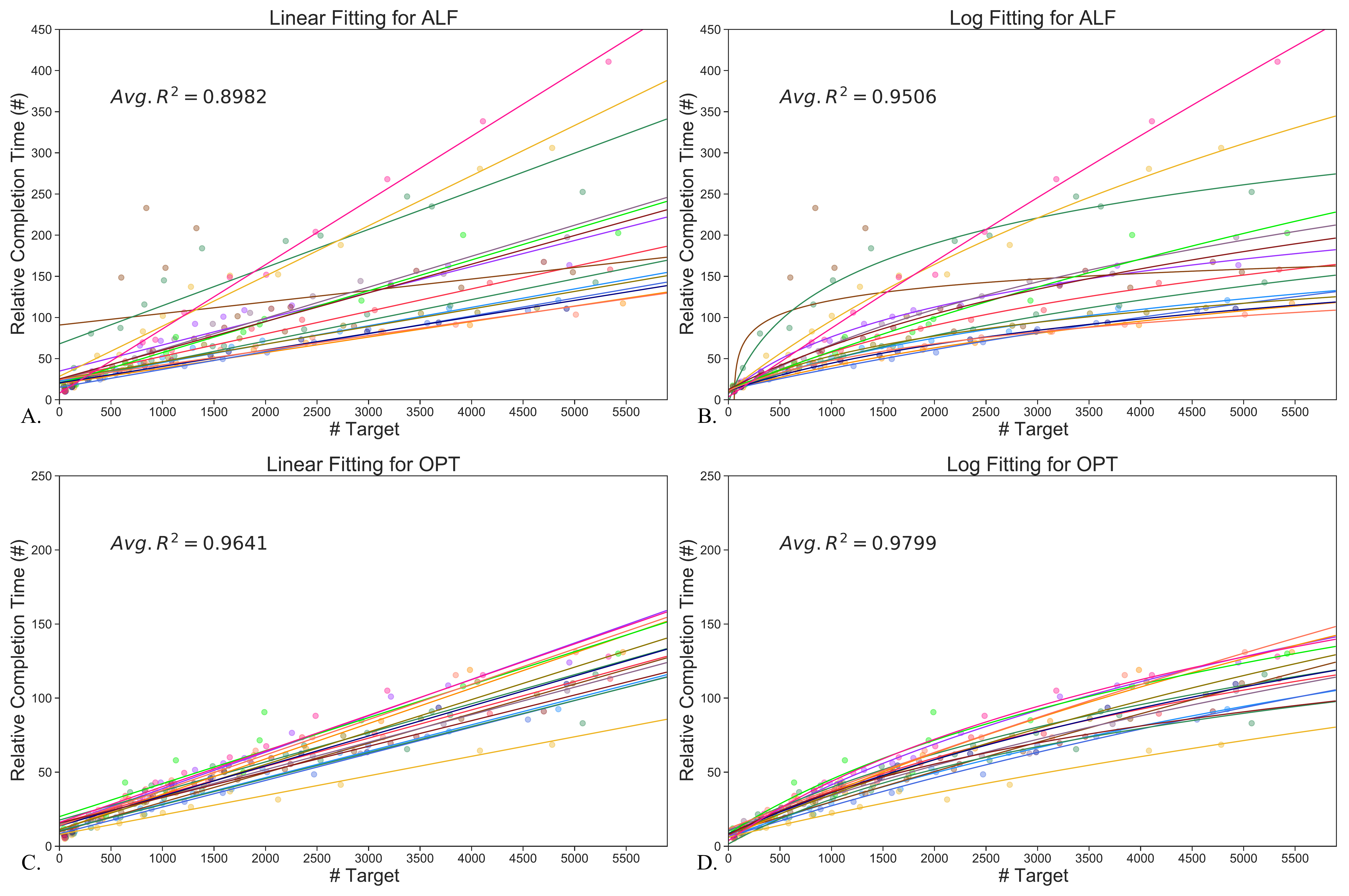}
\begin{center}
    Figure S3: The changing trends of relative completion time to form the 16 shapes in Table S1 as the shape scale grows, via ALF (our approach) and OPT-D, respectively.
\end{center}
\label{fig:iter-scale}
\end{figure}

The complete experimental results for evaluating scalability are presented in Figure S3 and S4:
\begin{itemize}
    \item[1.] Figure S3 shows the changing trends of relative completion time to form the 16 shapes in Table S1 as the shape scale grows, via ALF (our approach) and OPT-D, respectively. Two forms of fitting function (a linear function: $y=a\cdot n+b$, and a log function: $y=a\cdot log(b\cdot n+c)$, where $n$ is the independent variable, denoting the shape scale) are investigated.
    \item[2.] Figure S4 shows the changing trends of absolute completion time to form the 16 shapes in Table S1 as the shape scale grows, via ALF-1T (1-thread), ALF-16T (16-threads), and OPT-D, respectively. OPT-D is fitted by the function form of $y=a{\cdot n}^3+b{\cdot n}^2+c\cdot n+d$, and ALF is fitted by the function form of $y=\left(a\cdot n^2+b\cdot n+c\right)\cdot\left(d\cdot n+e\right)+f$.
\end{itemize}

The results show that,
\begin{itemize}
    \item[1.] For relative completion time, both methods achieve a $log(n)$-likely increase as the shape scale grows. Specifically, in the \emph{log} fitting, ALF and OPT-D achieve the $R^2$ of 0.9506 and 0.9799 on average in 16 shapes, respectively, better than the $R^2$ of 0.8982 and 0.9641 in \emph{linear} fitting.
    \item[2.] For absolute completion time, OPT-D achieves the $R^2>$0.99 (0.9996 on average) when fitting each shape’s experiment data by the $n^3$-form function, and ALF achieves the $R^2>$0.91 (0.9829 on average) when fitting each shape’s experiment data to the $n^2log\left(n\right)$-form function. Furthermore, ALF can be easily accelerated through parallelization, and the speedup of ALF-16T is 12.96.
\end{itemize}

\begin{figure}[h!]
\centering
\includegraphics[height=\dimexpr\pagegoal-\pagetotal\relax,width=\textwidth,keepaspectratio]{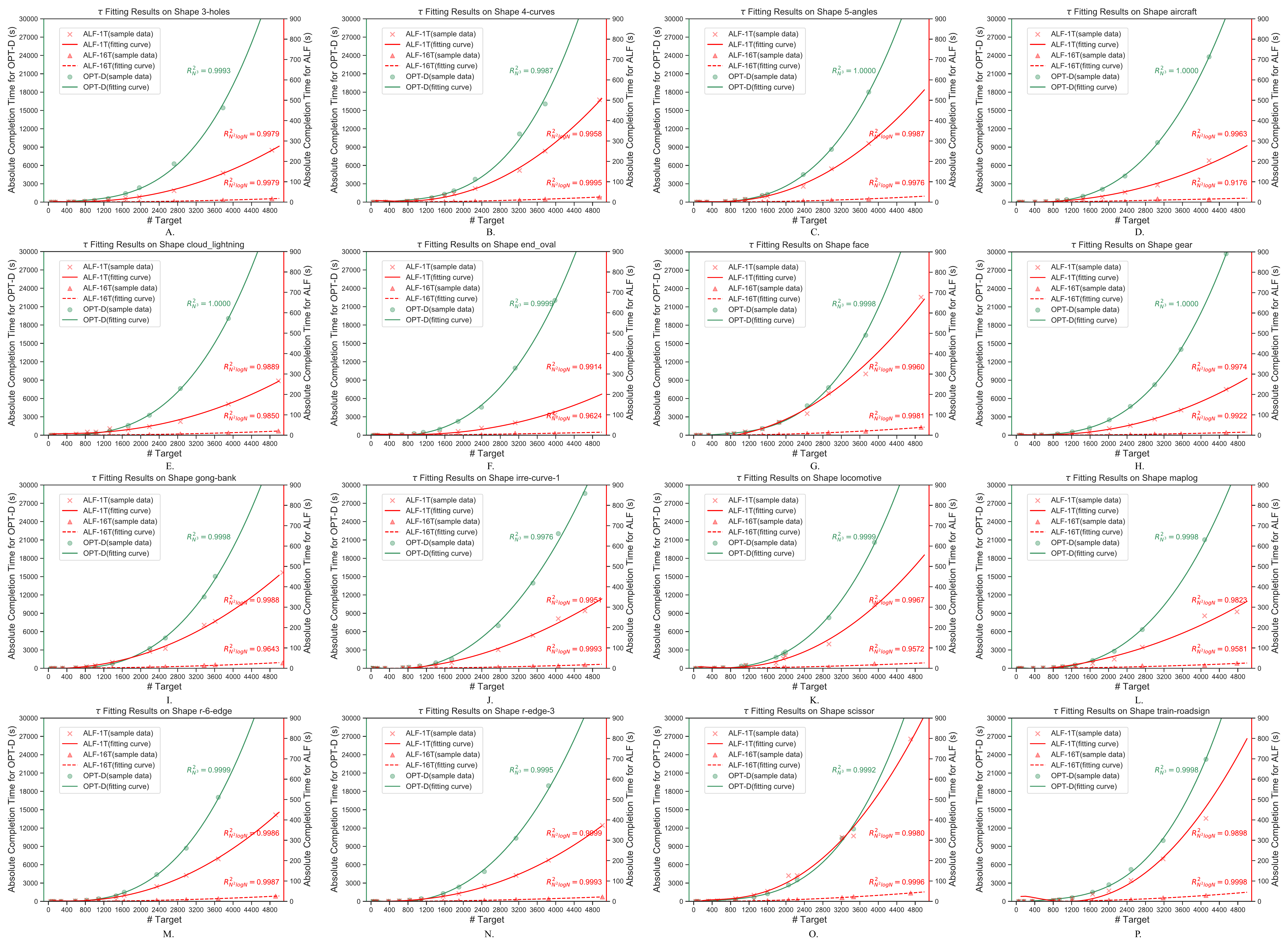}
\begin{center}
    Figure S4: The changing trends of absolute completion time to form each of the 16 shapes in Table S1 as the shape scale grows, via ALF-1T (1-thread), ALF-16T (16-threads), and OPT-D, respectively.
\end{center}
\label{fig:ab-time}
\end{figure}

In summary, compared to the state-of-the-art centralized distance-optimal algorithm OPT-D, ALF exhibits a $n^3$ to $n^2log\left(n\right)$ decrease in the absolute completion time of self-assembly tasks with respect to task scale $n$, and can be easily accelerated through parallelization, manifesting a good scalability.

\subsubsection{Stability}
The complete results for evaluating stability are presented by Table S3. The results show that ALF achieves normalized $\sigma(\rho)$=0.00034, $\sigma(t)$=0.04410, and $\sigma(\tau)$=0.00033 for the entire shape set in 156$\times$3$\times$50 experiments (156 shapes, 3 environment scales, and 50 repeated experiments), manifesting a high stability.

In addition, Figure S5 shows that the relative completion time of ALF and OPT-D are highly correlated: given a target shape $S$, and $N$ repeated experiments for an algorithm (ALF/OPT-D) to form $S$, the ratio of  $\hat{t}(S,ALF,N)$ over $\hat{t}(S,OPT-D,N)$, calculated by $\hat{r}(S,N)=\frac{\hat{t}(S,ALF,N)}{\hat{t}(S,OPT-D,N)}$, is relatively stable. Specifically, for most of the shapes (127 out of 156), the $\hat{r}(S,N)$ values keep relatively stable when the shape scale increases regardless the shape’s type (Figure S5.A); and most $\hat{r}(S,N)$ values are around 1.5 and do not exceed 3.6 in all experiments (Figure S5.B). Since the relative completion time of OPT-D has a theoretical upper bound (see Section 3.2), the highly correlated relation between the relative completion time of ALF and OPT-D indicates that ALF may also possess a similar property in the statistical sense.

\begin{figure}[h!]
\centering
\includegraphics[width=\textwidth,keepaspectratio]{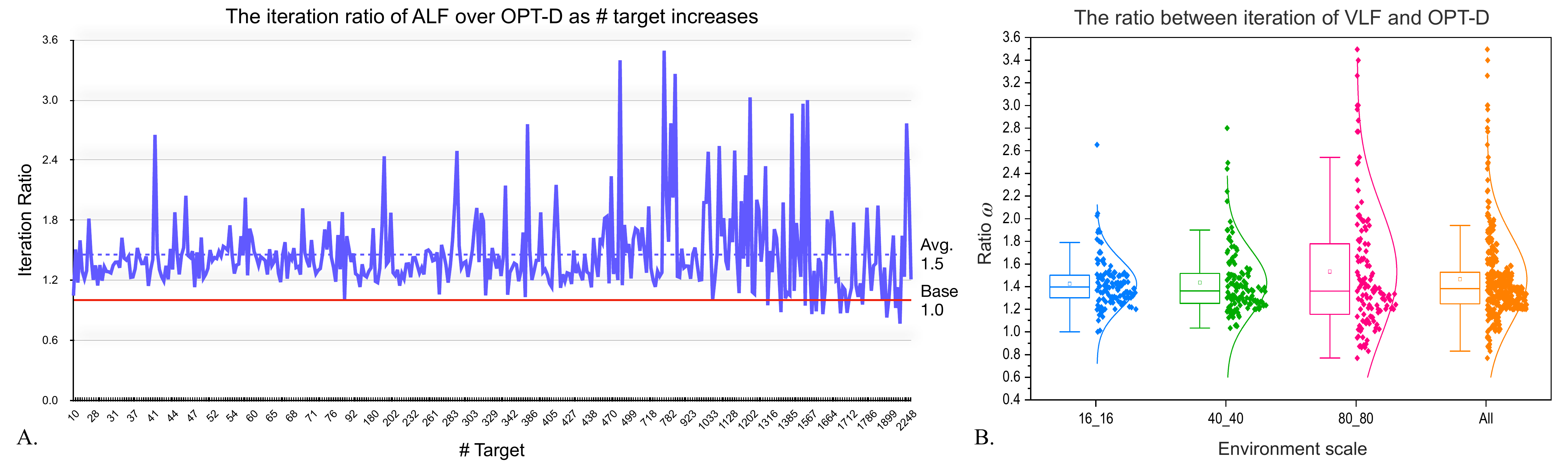}
\begin{center}
    Figure S5: The ratio of ALF’s relative completion time over that of OPT in 129 shapes with different shape/environment scales.
\end{center}
\label{fig:stable}
\end{figure}

\subsubsection{Discussion}
In experiments, we observed that HUN, OPT-D, DUD, and E-F have some weaknesses, resulting in their ineffective solutions to the self-assembly problem in grid environments. In the following, we elaborate on these weaknesses and analyze possible causes of these weaknesses.

Two weaknesses are observed in HUN:
\begin{itemize}
    \item[1.] 	The success rate of HUN decreases as the shape scale grows. HUN successfully formed 98.07\% shapes in 16$\times$16 environments, whereas in 40$\times$40 and 80$\times$80 environments, only 3.21\% and 0\% shapes were successfully formed by HUN. One possible cause of this weakness is that, when assigning target grids to agents based on shortest distances, HUN ignores the potential conflicts between paths assigned to agents, which are more likely to appear as the number of agents increases.
	\item[2.] Traffic jams often occur in HUN, resulting in low efficiency. Specifically, a traffic jam is a special kind of path conflict between agents, which occurs when agents reach their target grids located at the shape boundary earlier than those agents whose target grids located at the shape’s inner area, and thus prevent these agents from entering the shape. One possible cause of this weakness is that HUN does not take account of the temporal relation between agent movements during path planning. Figure S6 shows an example of traffic jam in HUN.
\end{itemize}

One weakness is observed in OPT-D: the absolute completion time of OPT-D is extremely high. The cause is that the two activities of agent-grid assignment and path-vertex ordering in OPT-D both have a high computational complexity of $O\left(n^3\right)$ for a self-assembly task with $n$ agents. Accordingly, the time efficiency of OPT-D could be improved from two points: using a distributed agent-grid assignment algorithm \cite{Chopra17} to improve parallelism; replacing global path-vertex ordering with a lightweight local priority negotiation protocol \cite{Wang20}.

Two weaknesses are observed in DUD: high frequency of agent collisions, and low completion quality. One possible cause for the two weaknesses is that the control strategy of DUD does not suitable for self-assembly tasks in grid environments. Specifically, when an agent enters the target shape, the attractive force will turn to 0, and the agent will keep moving along the same direction until some other agents appear in its neighborhood, making the agent changes its moving direction. For agents inside the target shape, this strategy leads to the phenomena of oscillation, i.e., each agent moves back and forth around its target grid. This strategy is suitable for self-assembly in continuous environments with sparse target distribution. But in the discrete grid environments with dense target distribution, the oscillation will cause an increase in the number of overlapping agents within the target shape, and the position-correcting activity in DUD cannot separate those overlapping agents correctly, resulting in a low completion quality.

\begin{figure}[h!]
\centering
\includegraphics[width=0.72\textwidth,keepaspectratio]{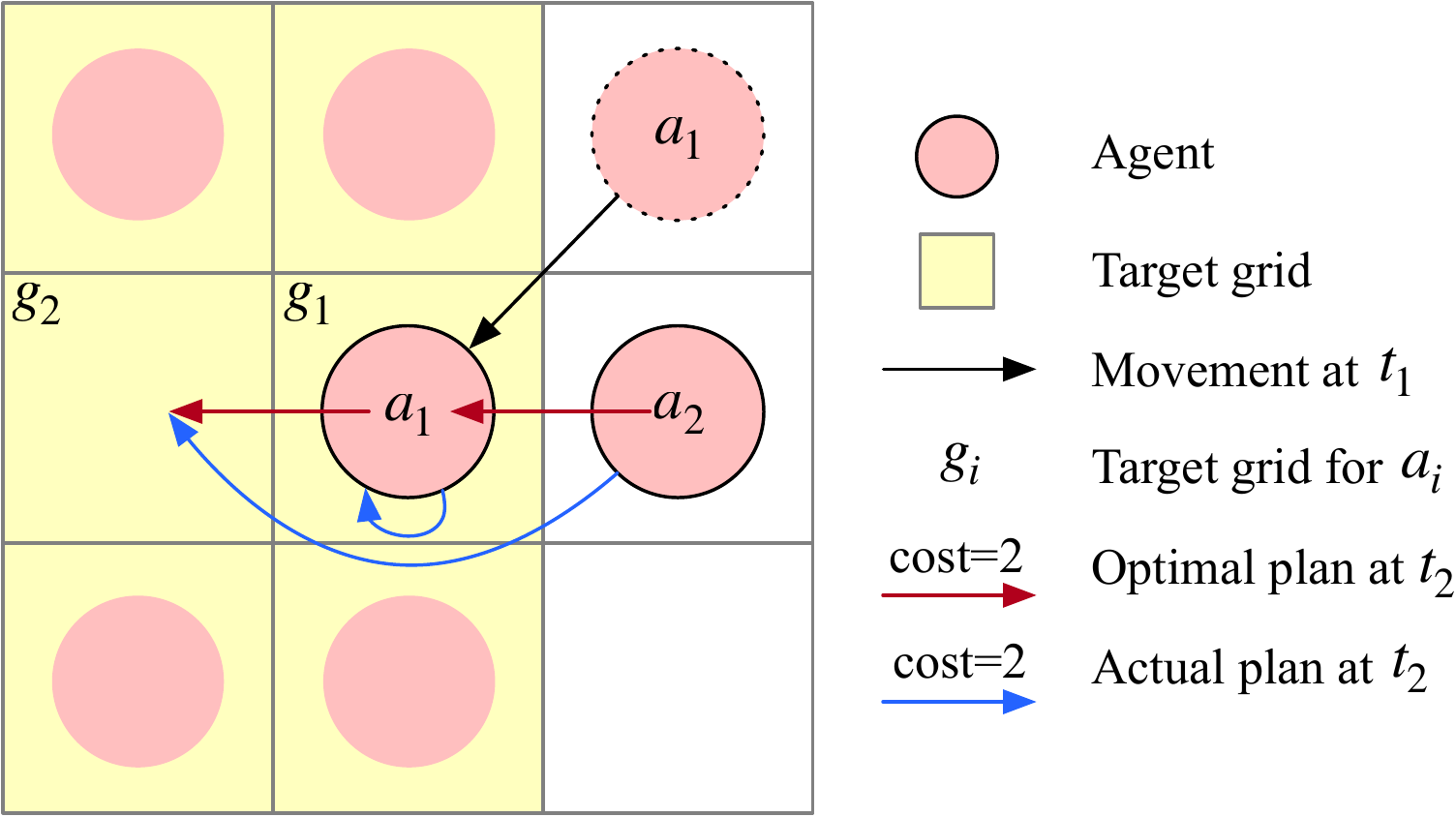}
\begin{center}\justifying
    Figure S6: An example of traffic jam in HUN. At time $t_1$, it is observed that agent $a_1$ moves first to its target grid $g_1$ and blocks the way of agent $a_2$ towards its target grid $g_2$, so re-planning is required before the next time step $t_2$. The optimal plan is that the $a_1$ and $a_2$ can swap their target grids and move left by one grid together with a total cost of 2. However, since HUN allocates goals directed by the minimal travel distance without considering path conflicts, so the actual plan may still be that $a_1$ stays at $g_1$ and $a_1$ moves to $g_2$, which has the same distance cost as the optimal plan but is impracticable.
\end{center}
\label{fig:HUN-bad-case}
\end{figure}

Two weaknesses are observed in E-F:
\begin{itemize}
    \item[1.] In general, the efficiency of E-F is extremely low. The cause is that the edge-following strategy greatly increases an agent’s travel distance from its initial position to its destination, and also greatly decreases the system parallelism because at any time only those agents that locate at the swarm’s boundary can move. 
    \item[2.] For shapes with holes, both the success rate and the completion quality and of E-F are low. When E-F terminates in a self-assembly task of a target shape with holes, there are usually many unoccupied areas around the holes within the formed shape; in extreme cases, E-F even never terminates. The cause is that the agent stop condition of E-F does not suitable for shapes with holes. In E-F, once entering the shape, an agent will stop moving when one of two conditions is satisfied: 1. the agent is about to move out of the shape; 2. the agent is next to a stopped agent with the same or greater gradient. From the second stop condition, the following property can be induced: an agent will stop moving as long as it connects two stopped agents with different gradients (such a scenario usually happens when the agent moves along a hole in the shape, even there are still unoccupied grids around the hole). The reason is that if a moving agent connects two stopped neighbors with gradients of $x$ and $y$ satisfying $x\neq y$, then the moving agent’s gradient will be updated to $min\left(x+1,y+1\right)$; since $x\neq y$, so $min\left(x+1,y+1\right)\le max(x,y)$, which means the agent reaches the second stop condition. The updated gradient will further propagate through connected agents, causing subsequent agents to stop moving earlier and thus resulting in many unoccupied grids. When an agent cannot enter the shape, it will never reach any stop condition and thus keep moving around the swarm's outer edge. An example is shown in Figure S7. 
\end{itemize}

\begin{figure}[th!]
\centering
\includegraphics[width=\textwidth,keepaspectratio]{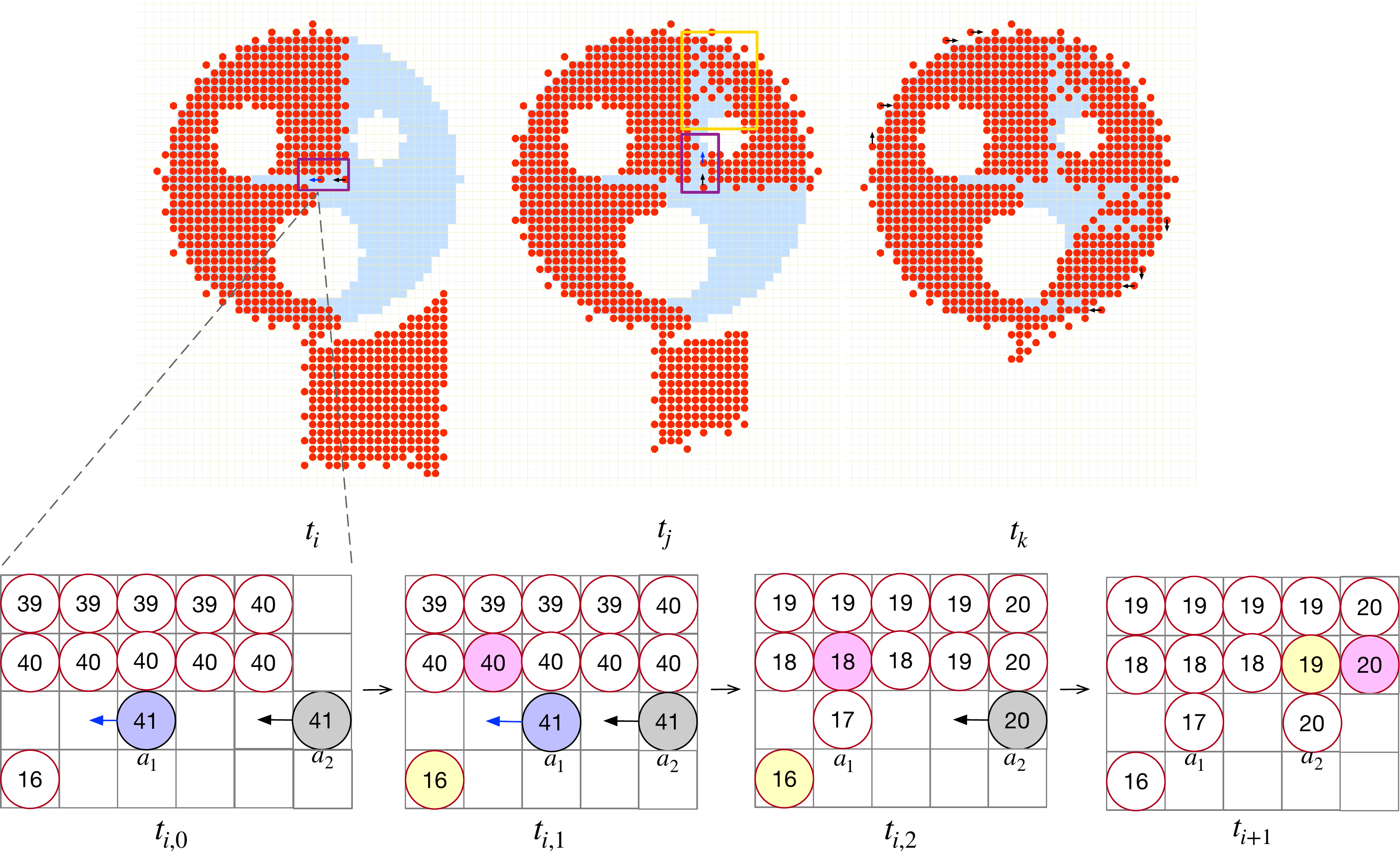}
\begin{center}\justifying
    Figure S7: An example of forming a shape with holes by E-F. At time $t_i$, three micro-steps ( $t_{i,0}$, $t_{i,1}$, and $t_{i,2}$) are observed: at $t_{i,0}$, the gradients of $a_1$ and $a_2$ are both 41, and both agents are planning to move left; at $t_{i,1}$, $a_2$ moves left, and its movement doesn’t trigger the update of gradients or any stop condition, so $a_2$ plans to keep moving left at next time step $t_{i+1}$; at $t_{i,2}$, $a_1$ moves left and connects the upper (pink) and the below (yellow) neighbors, with gradients of 40 and 16, respectively, which results in changes of agents’ gradients (in particular, the gradient of $a_1$ is changed to 17, and the gradient of its upper neighbor is changed accordingly to 18) and triggers the stop condition of $a_1$, leading to an unoccupied area at the left of $a_1$. At time $t_{i+1}$, agent $a_2$ moves left and triggers the stop condition of itself, leaving an unoccupied grid between $a_1$ and $a_2$. Subsequent agents will also stop earlier like $a_2$ due to the recalculation of gradients. At time $t_j$, many unoccupied areas appear in the yellow rectangle, and a similar scenario of $t_i$ occurs again in the purple rectangle, resulting in more unoccupied areas. At time $t_k$, many agents keep moving around the outer edge of the shape since they have no chance to enter the shape.
\end{center}
\label{fig:E-F-bad-case}
\end{figure}

\subsection{Parameter Analysis}
Our approach has four adjustable parameters: $\mathcal{W}\in[0,1]$ that controls the time of policy changing, $\mathit{flag} \in\{true,\,false\}$ that determines whether agents are allowed to move out of the target shape after entering, $\gamma\in[0,1]$ that indicates the probability of choosing actions that are worse than staying still, and f that represents one of the 9 types of the distance-discount function (see section 3.5). In order to investigate the effect of different parameter values on ALF’s performance, we select a subset of experiments on the two shapes of ``3-holes'' and ``r-6-edge'' from the experiments mentioned in section 3.5. In particular, for each of the four parameters, we fix other parameters to their values in the best parameter combination (i.e., $\mathcal{W}=0.15$, $\mathit{flag}=\mathit{True}$, $\gamma=0.2$, and $f=\mathit{type}$-6), change the parameter’s value (12 values of  $\mathcal{W}$, 6 values of $\gamma$, 2 values of $\mathit{flag}$, and 9 values of $f$), and observe ALF’s relative completion times on different values. The results are shown in Figure S8.

For parameter $\mathcal{W}$, it is observed that: when $\mathcal{W}=0$, the relative completion time is relatively high; as $\mathcal{W}$ increases to 0.05, the relative completion time decreases rapidly; after that, as $\mathcal{W}$ increases, the relative completion time decreases slowly, and when $\mathcal{W}=0.15$, the relative completion time achieves the minimal value; after that, as $\mathcal{W}$ increases further, the relative completion time also increase slowly. As a result, to achieve a shorter relative completion time, it is better to set $\mathcal{W}>0$.
	 
For parameter $\gamma$, it is observed that: when $\gamma=0$, the relative completion time of the algorithm is relatively low; when $\gamma=0.2$, the relative completion time achieves the minimal value; as $\gamma$ increases from 0.2 to 0.6, the relative completion time increases slowly; and as $\gamma$ increases further, the relative completion time increases rapidly. As a result, to achieve a shorter relative completion time, it is better to set $\gamma<0.5$.
	
For parameter $f$, it is observed that:
\begin{itemize}
    \item[1.] For \emph{linear} and \emph{inverse} light-intensity distance-discount functions, \emph{Chebyshev} distance measurement function performs better than other two distance measurement functions. For \emph{square-inverse} light-intensity distance-discount function, \emph{European} distance measurement function performs best.
    \item[2.] For all distance measurement functions, \emph{linear} light-intensity distance-discount function performs worse than the other two distance-discount functions, and the performance of \emph{inverse} and \emph{square-inverse} distance-discount functions shows little difference.
    \item[3.] The \textit{type}-6 $f$ function achieves the best performance on both shapes;
\end{itemize} 
As a result, to achieve a shorter relative completion time, it is better to set $f$ to \textit{type}-6 (the combination of \emph{inverse} light-intensity distance-discount function and \emph{Chebyshev} distance measurement function) or \textit{type}-8 (the combination of \emph{square-inverse} function and \emph{European} function).

For parameter $\mathit{flag}$, it is observed that:
\begin{itemize}
    \item[1.] Both shapes are formed faster when setting $\mathit{flag}$ as $True$ than $False$;
    \item[2.] The change of $\mathit{flag}$ shows much greater effects on the relative completion time when forming 3-holes than r-6-edge;
\end{itemize}
As a result, to achieve a shorter relative completion time, it is better to set $\mathit{flag}$ as $True$.

\begin{figure}[h!]
\centering
\includegraphics[width=\textwidth,keepaspectratio]{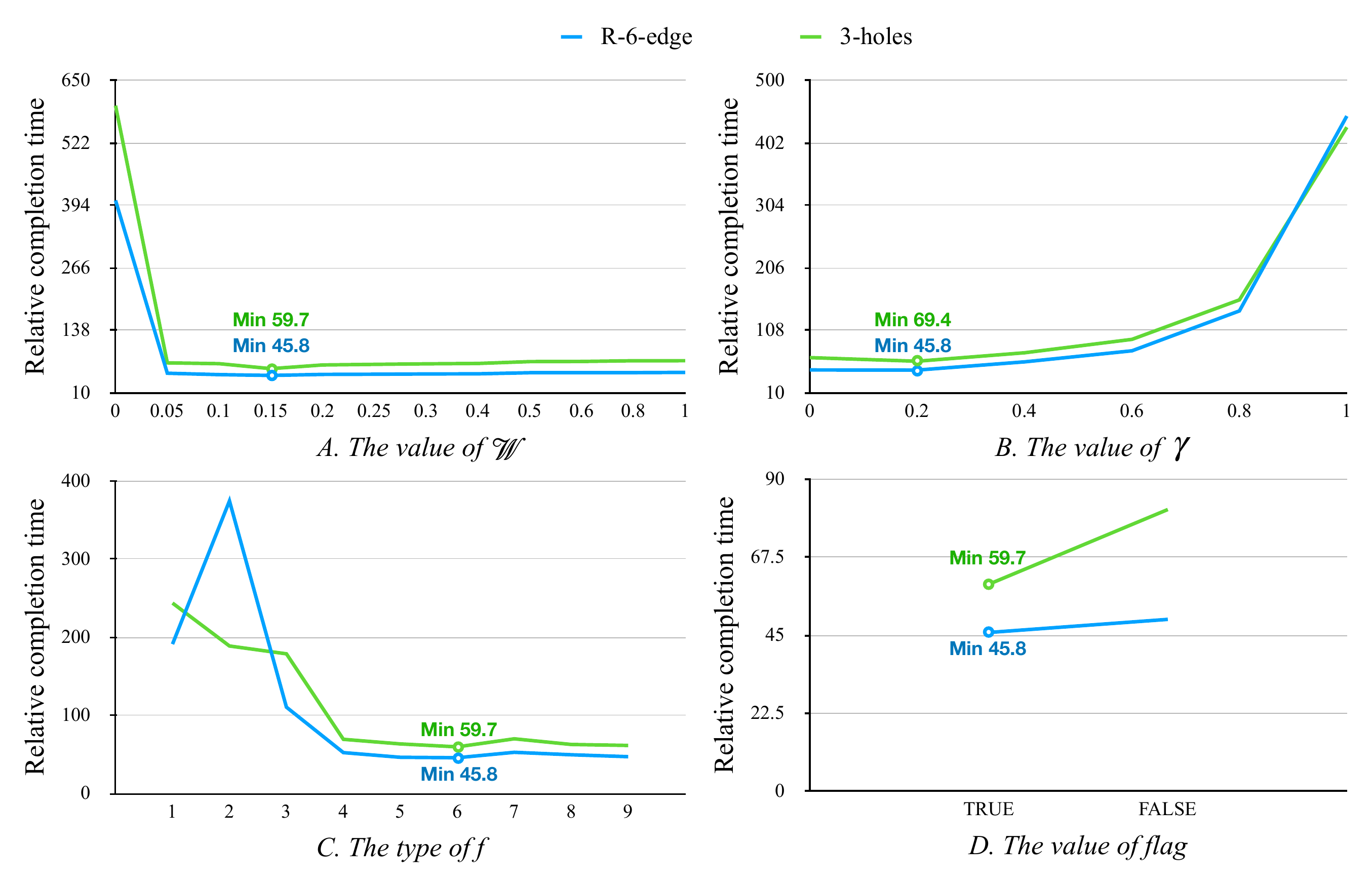}
\begin{center}
    Figure S8: The effect of different parameter values on the efficiency of ALF.
\end{center}
\label{fig:param}
\end{figure}


\begin{thebibliography}{100}

\bibitem{Whi02}
G.~M. Whitesides, B.~Grzybowski, Self-assembly at all scales. {\it Science\/} {\bf 295}, 2418 (2002).

\bibitem{Grzybowski09}
B.~A. Grzybowski, C.~E. Wilmer, J.~Kim, K.~P. Browne, K.~J.~M. Bishop, Self-assembly: from crystals to cells. {\it
  Soft Matter\/} {\bf 5}, 1110 (2009).

\bibitem{Estroff04}
L.~A. Estroff, A.~D. Hamilton, Water gelation by small organic molecules. {\it Chemical Reviews\/} {\bf 104}, 1201 (2004).
  PMID: 15008620.

\bibitem{Marsh15}
J.~A. Marsh, S.~A. Teichmann, Structure, dynamics, assembly, and evolution of protein complexes. {\it Annual Review of Biochemistry\/} {\bf 84},
  551 (2015). PMID: 25494300.

\bibitem{Weijer09}
C.~J. Weijer, Collective cell migration in development. {\it Journal of Cell Science\/} {\bf 122}, 3215 (2009).

\bibitem{Mehes14}
E.~Méhes, T.~Vicsek, Collective motion of cells: from experiments to models. {\it Integrative Biology\/} {\bf 6}, 831 (2014).

\bibitem{camazine03}
S.~Camazine, {\it et~al.\/}, {\it Self-organization in biological systems\/},
  vol.~7 (Princeton university press, 2003).

\bibitem{mlot11}
N.~J. Mlot, C.~A. Tovey, D.~L. Hu, Fire ants self-assemble into waterproof rafts to survive floods. {\it Proceedings of the National Academy of
  Sciences\/} {\bf 108}, 7669 (2011).

\bibitem{Finn07}
A.~{Finn}, K.~{Kabacinski}, S.~P. {Drake}, Design challenges for an autonomous cooperative of UAVs. {\it 2007 Information, Decision and
  Control\/} (2007), pp. 160--169.

\bibitem{KevinZ18}
K.~Z.~Y. Ang, {\it et~al.\/}, High-precision multi-UAV teaming for the first outdoor night show in Singapore. {\it Unmanned Syst.\/} {\bf 6}, 39 (2018).

\bibitem{Viriyasitavat12}
W.~{Viriyasitavat}, O.~K. {Tonguz}, Priority Management of Emergency Vehicles at Intersections Using Self-Organized Traffic Control. {\it 2012 IEEE Vehicular Technology
  Conference (VTC Fall)\/} (2012), pp. 1--4.

\bibitem{Daniel18}
D.~Str{\"{o}}mbom, A.~Dussutour, Self-organized traffic via priority rules in leaf-cutting ants. {\it PLoS Computational Biology\/} {\bf 14}
  (2018).

\bibitem{rubenstein14}
M.~Rubenstein, A.~Cornejo, R.~Nagpal, Programmable self-assembly in a thousand-robot swarm. {\it Science\/} {\bf 345}, 795 (2014).

\bibitem{tucci18}
T.~Tucci, B.~Piranda, J.~Bourgeois, A distributed self-assembly planning algorithm for modular robots. {\it Proceedings of the 17th International
  Conference on Autonomous Agents and MultiAgent Systems\/}, pp. 550--558.

\bibitem{JYu13}
J.~{Yu}, S.~M. {LaValle}, Shortest path set induced vertex ordering and its application to distributed distance optimal formation path planning and control on graphs. {\it 52nd IEEE Conference on Decision and Control\/}
  (2013), pp. 2775--2780.

\bibitem{alonso11}
J.~Alonso-Mora, A.~Breitenmoser, M.~Rufli, R.~Siegwart, P.~Beardsley, Multi-robot system for artistic pattern formation. {\it 2011
  IEEE international conference on robotics and automation\/} (IEEE, 2011), pp.
  4512--4517.

\bibitem{sabattini09}
L.~Sabattini, C.~Secchi, C.~Fantuzzi, Potential based control strategy for arbitrary shape formations of mobile robots. {\it 2009 IEEE/RSJ International
  Conference on Intelligent Robots and Systems\/} (IEEE, 2009), pp. 3762--3767.

\bibitem{chiang15}
H.-T. Chiang, N.~Malone, K.~Lesser, M.~Oishi, L.~Tapia, Path-guided artificial potential fields with stochastic reachable sets for motion planning in highly dynamic environments. {\it 2015 IEEE
  International Conference on Robotics and Automation (ICRA)\/} (IEEE, 2015),
  pp. 2347--2354.

\bibitem{falomir18}
E.~Falomir, S.~Chaumette, G.~Guerrini, A Mobility model based on improved artificial potential fields for swarms of UAVs. {\it 2018 IEEE/RSJ International
  Conference on Intelligent Robots and Systems (IROS)\/} (IEEE, 2018), pp.
  8499--8504.

\bibitem{bi18}
Q.~Bi, Y.~Huang, A self-organized shape formation method for swarm controlling. {\it 2018 37th Chinese Control Conference (CCC)\/} (IEEE,
  2018), pp. 7205--7209.

\bibitem{wolf04}
J.~Wolf, P.~Robinson, J.~Davies, Vector field path planning and control of an autonomous robot in a dynamic environment. {\it the Proceedings of the 2004 FIRA Robot
  World Congress (Paper 151)\/} (2004).

\bibitem{gayle09}
R.~Gayle, W.~Moss, M.~C. Lin, D.~Manocha, Multi-robot coordination using generalized social potential fields. {\it 2009 IEEE International
  Conference on Robotics and Automation\/} (IEEE, 2009), pp. 106--113.

\bibitem{sabattini11}
L.~Sabattini, C.~Secchi, C.~Fantuzzi, Arbitrarily shaped formations of mobile robots: artificial potential fields and coordinate transformation. {\it Autonomous Robots\/} {\bf 30}, 385
  (2011).

\bibitem{theraulaz99}
G.~Theraulaz, E.~Bonabeau, A brief history of stigmergy. {\it Artificial Life\/} {\bf 5}, 97 (1999).

\bibitem{zhang20}
W.~Zhang, H.~Mei, A constructive model for collective intelligence. {\it National Science Review\/}  {\bf 8}, 7 (2020).

\bibitem{Jekely2009}
G.~J{\'e}kely, {\it Philosophical Transactions of the Royal Society B:
  Biological Sciences\/} {\bf 364}, 2795 (2009).

\bibitem{Dalessandro11}
L.~Dalessandro, D.~Dice, M.~Scott, N.~Shavit, M.~Spear, Transactional mutex locks. {\it Euro-Par 2010 -
  Parallel Processing\/}, P.~D'Ambra, M.~Guarracino, D.~Talia, eds. (Springer
  Berlin Heidelberg, Berlin, Heidelberg, 2010), pp. 2--13.
  
\bibitem{hcheng}
H.~Cheng, Q.~Zhu, Z.~Liu, T.~Xu, L.~Lin, Decentralized navigation of multiple agents based on orca and model predictive control. {\it 2017 IEEE/RSJ International Conference on Intelligent Robots and Systems (IROS)\/} (IEEE, 2017), pp. 3446-3451.

\bibitem{Chopra17}
S.~Chopra, G.~Notarstefano, M.~Rice, M.~Egerstedt, A distributed version of the Hungarian method for multirobot assignment. {\it IEEE Transactions on Robotics\/}, {\bf 33}, 932 (2017).

\bibitem{Wang20}
H.~Wang, M.~Rubenstein, Shape formation on homogeneous swarms using local task swapping. {\it IEEE Transactions on Robotics\/}, {\bf 36}, 597 (2020).

\end{thebibliography}
\end{document}